%% file: ms.tex
\documentclass[11pt,preprint]{aastex}
\usepackage{psfig}

\newcounter{species}
\def\ion#1#2{\setcounter{species}{#2}#1$\;${\sc\roman{species}}\relax}
\def\lion#1#2#3{
\setcounter{species}{#2}#1$\;${\sc\roman{species}}$\;\lambda${#3}\relax
}
\def\llion#1#2#3{
\setcounter{species}{#2}#1$\;${\sc\roman{species}}$\;\lambda\lambda${#3}\relax
}

\def\kms{\ifmmode{~{\rm km~s^{-1}}}\else{~km s$^{-1}$}\fi}
\def\lsim{\lower0.3em\hbox{$\,\buildrel <\over\sim\,$}}
\def\gsim{\lower0.3em\hbox{$\,\buildrel >\over\sim\,$}}
\def\hst{{\it HST}}



\begin{document}

\shorttitle{Variable UV Absorption Lines in Quasars}
\shortauthors{Wise et al.}

\title{Variability of Narrow, Associated Absorption Lines in
Moderate- and Low-Redshift Quasars}

\author{John H. Wise, Michael Eracleous, Jane C. Charlton, and Rajib
Ganguly\altaffilmark{1}}

\affil{Department of Astronomy and Astrophysics, 525 Davey Laboratory,
The Pennsylvania State University, University Park, PA 16802}

\email{jwise, mce, charlton, ganguly@astro.psu.edu}

\altaffiltext{1}{Current address: Space Telescope Science Institute,
3700 San Martin Drive, Baltimore, MD 21218; e-mail {\tt
ganguly@stsci.edu}.}


\begin{abstract}
We present the results of a search for variability in the equivalent
widths (EWs) of narrow, associated ($\vert\Delta v\vert \leq
5,000\kms$) absorption lines found in the UV spectra of $z \leq 1.5$
quasars.  The goal of this search was to use variability as a means of
identifying absorption lines arising in gas that is intrinsic to the
quasar central engine. We have compared archival \hst/FOS spectra of
quasars with recent spectra obtained as part of our own snapshot
survey of the same objects with STIS.  The intervals between
observations are 4--10 years.  We primarily focused on the \ion{C}{4}
absorption lines, although we also studied other lines available in
the same spectra (e.g., Ly$\alpha$, \ion{N}{5}, \ion{O}{6}).  Our main
result is that 4 out of 15 quasars, or 4 out of 19 associated
absorption systems, contained variable narrow absorption lines, which
are indicative of intrinsic absorption.  We conclude that a minimum of
21\% of the associated absorption-line systems are variable. Because
not all systems will have necessarily varied, this is a lower limit on
this fraction and it is consistent with previous estimates based on
variability, partial coverage analysis, or statistical arguments.  If
we interpret the upper limits on the variability time scale as upper
limits on the recombination time of the absorber, we constrain the
density of the absorber to be $n_e > 3000~{\rm cm}^{-3}$ and its
distance from the ionizing source to be $R\lsim 100$~pc.  Moreover, we
are now able to pick out specific intrinsic absorption-line systems to
be followed up with high-dispersion spectroscopy in order to constrain
the geometry, location, and physical conditions of the absorber. We
briefly illustrate how followup studies can yield such constraints by
means of a simulation.

\end{abstract}


\section{Introduction}

The spectra of quasars often show a plethora of absorption lines, some
of which arise in gas that is somehow associated with the quasar,
while others arise in systems that are unrelated to the quasar (e.g.,
galaxies along the line of sight at a much lower redshift). A recent
summary of the properties of intrinsic absorption lines (and of their
nomenclature) can be found in \citet{HaSa04}. The properties of
absorbing gas that is intimately associated with the quasar are of
particular interest in studies of quasar central engines because they
provide constraints and insights on the accretion process that powers
the central engine (such gaseous systems are referred to hereafter as
``intrinsic systems'' and the lines they produce as ``intrinsic
absorption lines'') . The predominance of {\it blueshifts} among
intrinsic absorption lines, suggests that the gas has the form of an
outflowing wind, which in turn makes it an important component of the
accretion process. More specifically, the wind can be an important
mechanism for extracting angular momentum from the accretion flow,
allowing accretion to proceed. Thus the study of intrinsic absorption
lines can provide information on how common winds are, as well as
estimates or limits on the mass outflow rates, which can serve as
constraints on models for the accretion flow. Studying the dynamics
and conditions in the absorber also allows us to explore its connection to
the broad-emission line gas.  In addition, intrinsic absorption lines
trace the evolution of quasar outflows, and presumably the cosmic
evolution in quasar fueling rate \citep[see, for example, the
discussion by][]{Ga_etal01}.

Before any detailed study of the physical conditions in intrinsic
absorption systems can be carried out, one must identify intrinsic
absorption lines with certainty, a task that can be rather difficult
in practice. In the case of {\it broad} absorption lines (BALs), whose
widths exceed 2,000\kms\ by definition and often exceed 10,000\kms\ in
practice \citep[e.g.,][]{Tu87,Tu88,We91}, a good case can be made that
the absorbing gas has the form of a fast accretion-disk wind
\citep[see, for example,][]{Mu95,Pro00}. Similarly, narrow absorption
lines (hereafter NALs; those whose widths are small enough that the
strong UV resonance doublets, such as \ion{C}{4}, \ion{Si}{4}, and
\ion{N}{5}, are resolved, i.e., FWHM$\lsim$300\kms), with $z_{\rm a}
\ll z_{\rm e}$ (where $z_{\rm a}$ and $z_{\rm e}$ are the absorption
emission redshifts, respectively), or more specifically with
velocities of 18,000\kms\ or more, relative to the quasar, are likely
to arise in gas that is distant from and not related to the quasar
\citep[see][]{We79a}. This is not guaranteed to always be the case,
however, since a number of NALs with $z_{\rm a} \ll z_{\rm e}$ have
been found to vary, which suggests that the absorber is closely
related to the quasar \citep*[e.g., in Q2343+125 and
PG~2302+029;][]{Ha97b,Jan02}. Moreover the statistical study of
\citet{Ri99} found that $dN/dz$ (the number of absorption-line systems
per object per unit redshift) in the range $1.6<z<3.5$ depends on
quasar optical and radio properties. More specifically, they find a
statistical excess of absorbers in the velocity range $-15,000\kms$ to
$-65,000\kms$ in optically luminous quasars ($M_{\rm V} < -27.0$),
relative to quasars of lower luminosities.

In the velocity range $\Delta v \lsim -18,000\kms$, $dN/dz$ is
significantly higher than the expected density of intervening
absorbers \citep{We79a}. In fact, at $\Delta v < -5,000\kms$ as many
as 2/3 of the absorption systems could be intrinsic to the quasars
\citep*[see, for example,][and references therein]{Ald94}. This excess
of absorbers could be attributed to a combination of galaxies in a
cluster surrounding the quasar and parcels of gas ejected from the
quasar in the form of a wind \citep{We79a,AWFJ87}.  The NALs of the
latter type are particularly interesting and deserving of detailed
study in the context of quasar central engines because they appear to
be ubiquitous in quasar spectra \citep[e.g.,][]{AWFJ87} and their
frequency appears to evolve with redshift \citep{Ga_etal01}. However,
judging which particular NALs arise in gas intrinsic to the quasar
is rather challenging.

This leads us to undertake this study, whose goal is to identify
intrinsic NALs in moderate- and low-redshift quasars ($z\leq 1.5$)
based on their variability. This is a relatively economical technique
discussed by \citet{BaSa97}.  We focus on absorption lines that are
within 5,000\kms\ of the quasar redshift; such absorption-line systems
are conventionally regarded as ``associated'' with the quasar,
following \citet{Fo86}.  Although many statistical studies have been
conducted to determine the frequency of intrinsic absorbers relative
to all NAL systems, only a handful of specific absorption systems have
been identified as truly intrinsic. Moreover, variability information
is sparse.  \citet{Ba97} estimated that 30\% of NALs are variable to
some unspecified level, while specific variable systems in
high-redshift quasars have been studied in detail by \cite{Ha95},
\citet{Ha97b}, and more recently by \citet{Na04}.

To search for variability among associated NALs in $z\leq 1.5$ quasars
we have carried out a snapshot survey with the Space Telescope Imaging
Spectrograph (STIS) on the {\it Hubble Space Telescope} (\hst). We
chose quasars which had already been observed with the \hst's Faint
Object Spectrograph (FOS) and were known to have associated
\ion{C}{4}, \ion{N}{5}, or \ion{O}{6} NALs.  The specific goals of
this survey were to: 
\\
(a) Determine the fraction of variable NALs among
the associated NALs observed in low-redshift quasars. This provides a
{\em lower limit} on the fraction of associated NALs that are truly
intrinsic, since not all intrinsic NALs will have varied while we
were observing them.  The results obtained for high-redshift quasars,
observed from the ground, need not apply at low redshifts, since the
frequency of associated NALs evolves with redshift \citep{Ga_etal01}.
\\
(b) Increase the number of confirmed intrinsic absorbers so that more
detailed followup studies can be carried out on {\it specific} systems.

We describe the observations and data analysis methods in \S2 while in
\S3 we present the results. In \S4 we discuss the implications of NAL
variability, we comment on individual objects, and suggest future,
followup work on NAL quasars.

\section{Observations and Data Analysis}\label{S_obs}

A sample of 15 quasars with associated NALs were surveyed for
variability primarily in the following UV absorption lines:
\llion{O}{6}{1032,1038}, \lion{H}{1}{1215}, \llion{N}{5}{1239,1243},
and \llion{C}{4}{1548,1551}, but also in the \llion{Si}{2}{1190,1193},
\lion{Si}{2}{1260}, and \lion{Si}{3}{1207} lines.  The journal of
observations is given in Table~\ref{tab_obs}, which also includes
information for both the first-epoch FOS observations and the
second-epoch STIS observations.  We have paid particular attention to
collecting accurate redshifts from the literature, since these are
important in determining the velocities of absorption lines relative
to the rest-frame of the quasar, and whether a specific absorption
system is blueshifted or redshifted relative to the quasar. Thus, we
have examined the original redshift reports for all of our target
quasars and considered their uncertainties before adopting the
reported values. In \S\ref{S_indiv}, we give a brief discussion of our
adopted redshift for specific objects.  In Table~\ref{tab_obs}, we
list the redshifts we have adopted along with their uncertainties,
where available, and the source of the information.  In cases where an
uncertainty is not quoted, we expect that the uncertainty is of order
a few in the last decimal place given. An uncertainty in the redshift,
$\delta z$, translates into an uncertainty in the velocity of an
absorption system relative to the quasar rest-frame of $\delta v\sim
c\;\delta z/(1+z) < c\;\delta z$. Redshift uncertainties of $\delta z
\lsim 10^{-4}$ are comparable to redshift errors arising from
heliocentric corrections; since the papers reporting the redshifts
rarely report the application of heliocentric corrections, an
additional uncertainty of up to 30\kms, may have to be taken into
account.

As Table~\ref{tab_obs} shows, not all desirable lines were observed in
all objects, although the \ion{C}{4} line (the primary target) was
covered in the spectra of most objects.  The first-epoch FOS spectra
of these objects (through the G190H and G270H gratings) were collected
in the early to mid 1990's and were kindly provided, fully and
uniformly reduced, by Sophia Kirhakos and Buell Jannuzi
\citep[see][]{Sc93}. A large fraction of them were part of the \hst\
Quasar Absorption Line Key Project (hereafter, KP). Part of the
reduction process was to correct the absolute wavelength scale; we
have checked these corrections by comparing the observed wavelengths
of securely-identified Galactic lines with their expected values and
found that the agreement to be better than 0.1~\AA\ half of the time
and never worse than 0.27~\AA.  In our followup snapshot program, the
quasars were observed with STIS using the G230L grating and a
$52\arcsec\times 0.\!\!\arcsec2$ slit.  This grating obtains a
resolution of $R \sim 1000$ and a dispersion of 1.58~\AA\ per pixel
with two pixels per resolution element \citep{Proffitt02}.  This
spectral resolution corresponds to a velocity resolution of
300--600\kms\ in the STIS G230L spectral range, compared to a
resolution of about 230\kms\ in the FOS spectra with four overlapping
pixels per resolution element \citep{Keyes95}.  The spectral range for
the STIS G230L is 1570--3180~\AA, whereas the FOS G190H and G270H
spectral ranges are 1573--2330~\AA\ and 2221--3301~\AA, respectively.
These gratings from different instruments conveniently overlap so that
a variability analysis can be conducted on the chosen objects.

\begin{figure} 
\centerline{\includegraphics[angle=-90,scale=0.55]{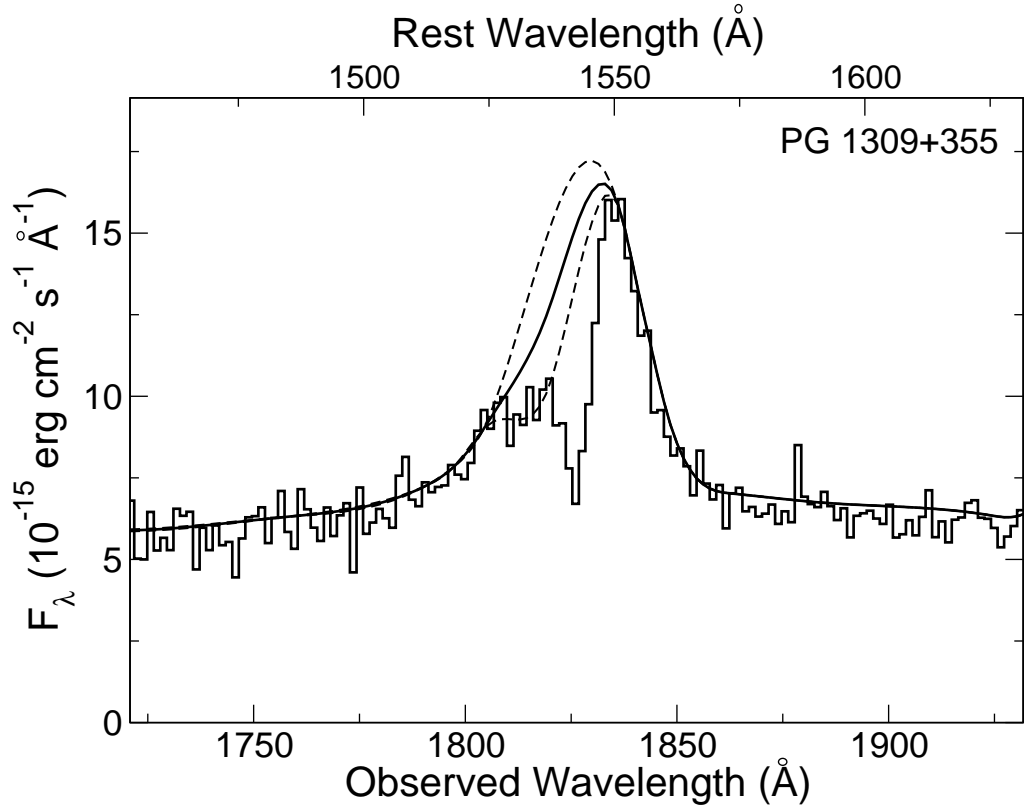}}
\caption{An example of extreme continuum fits to the STIS spectrum of
  a broad \ion{C}{4} emission line. We incorporate these uncertainties
  into our selection criteria of variable lines. An additional
  illustration using a FOS spectrum can be found in
  Figure~\ref{fig_pksfits}.
\label{fig_cont}}
\end{figure}

\begin{figure}  
\centerline{\includegraphics[scale=0.4,angle=90]{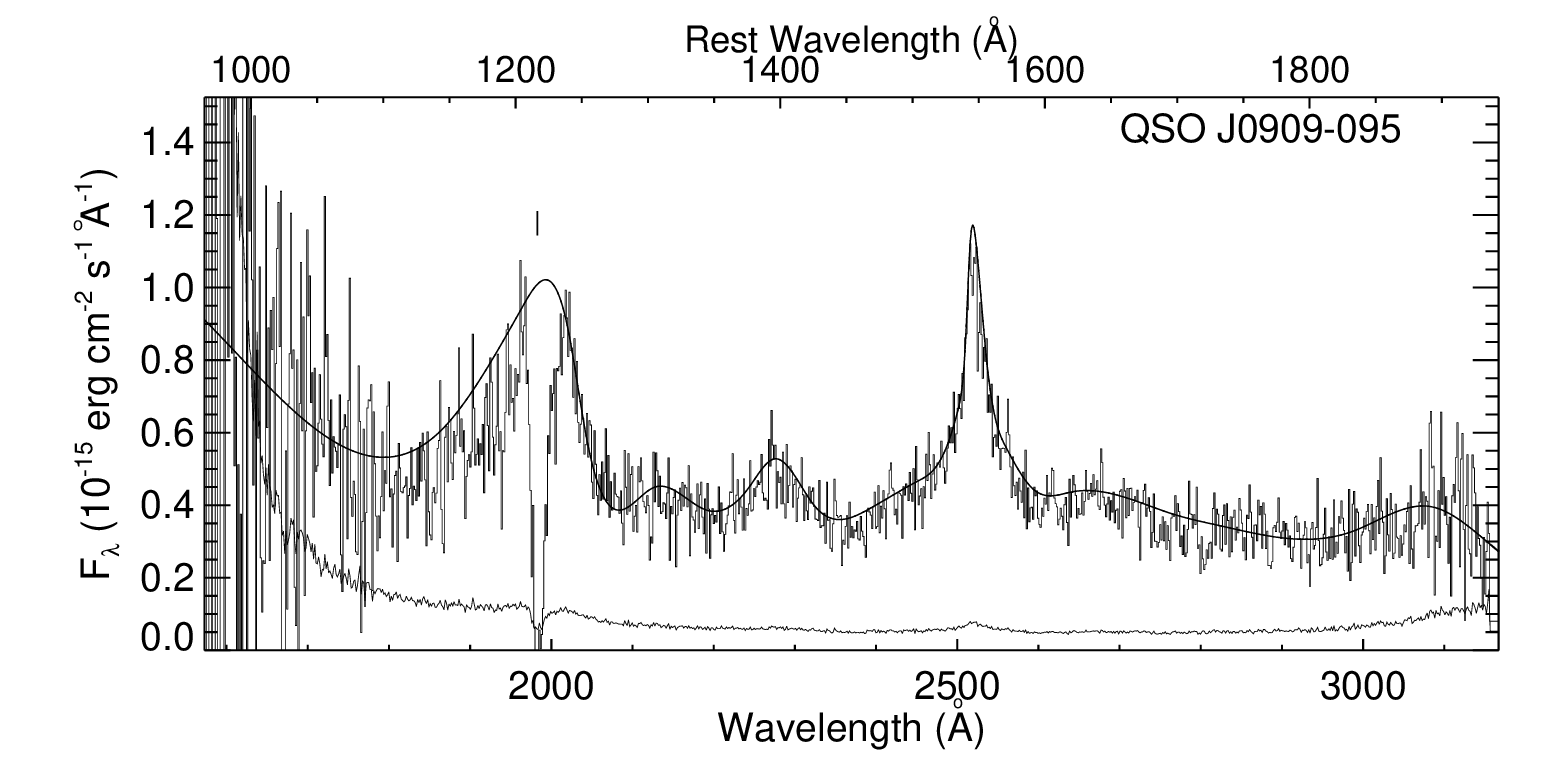}
            \includegraphics[scale=0.4,angle=90]{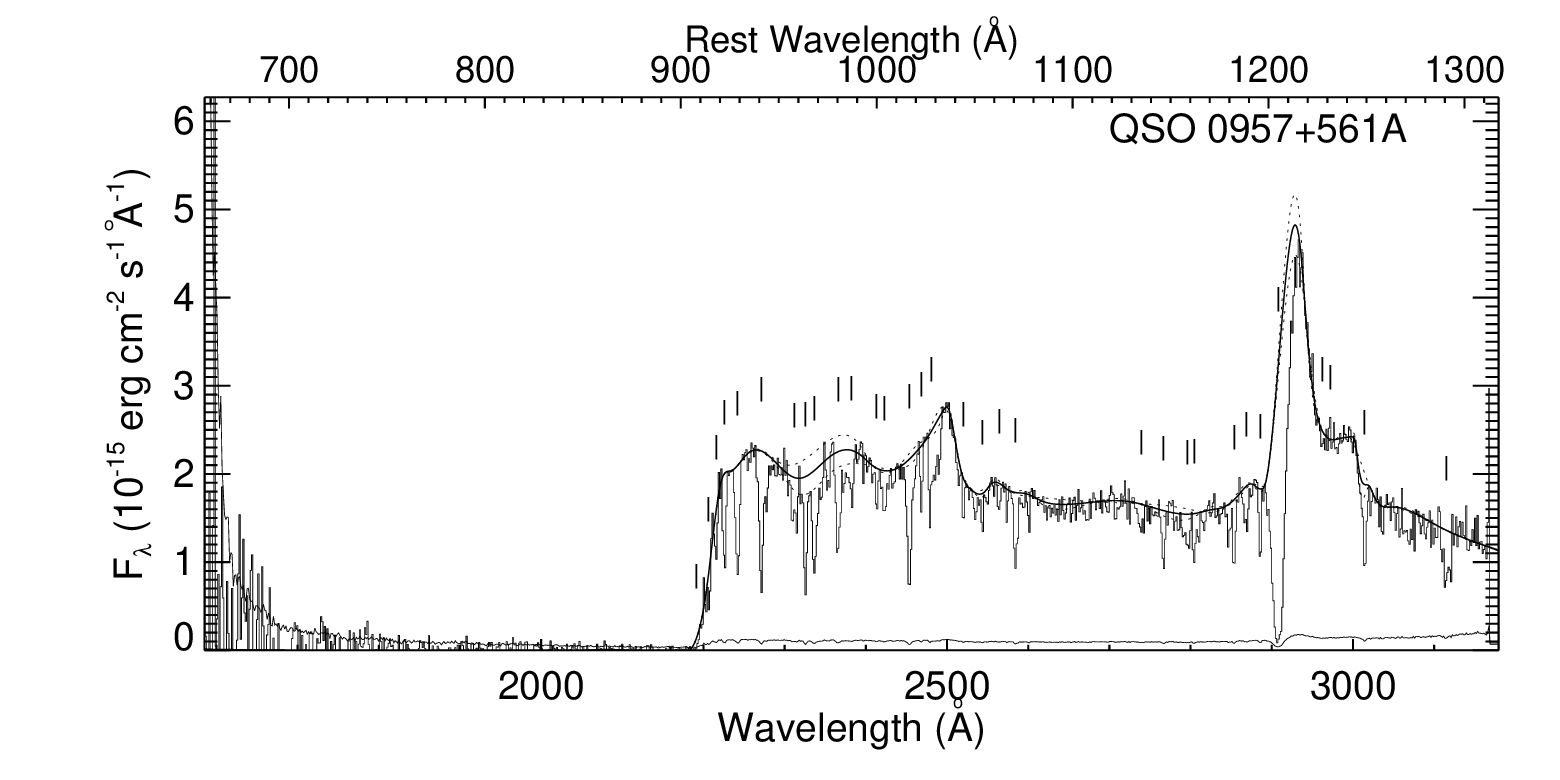}
            \includegraphics[scale=0.4,angle=90]{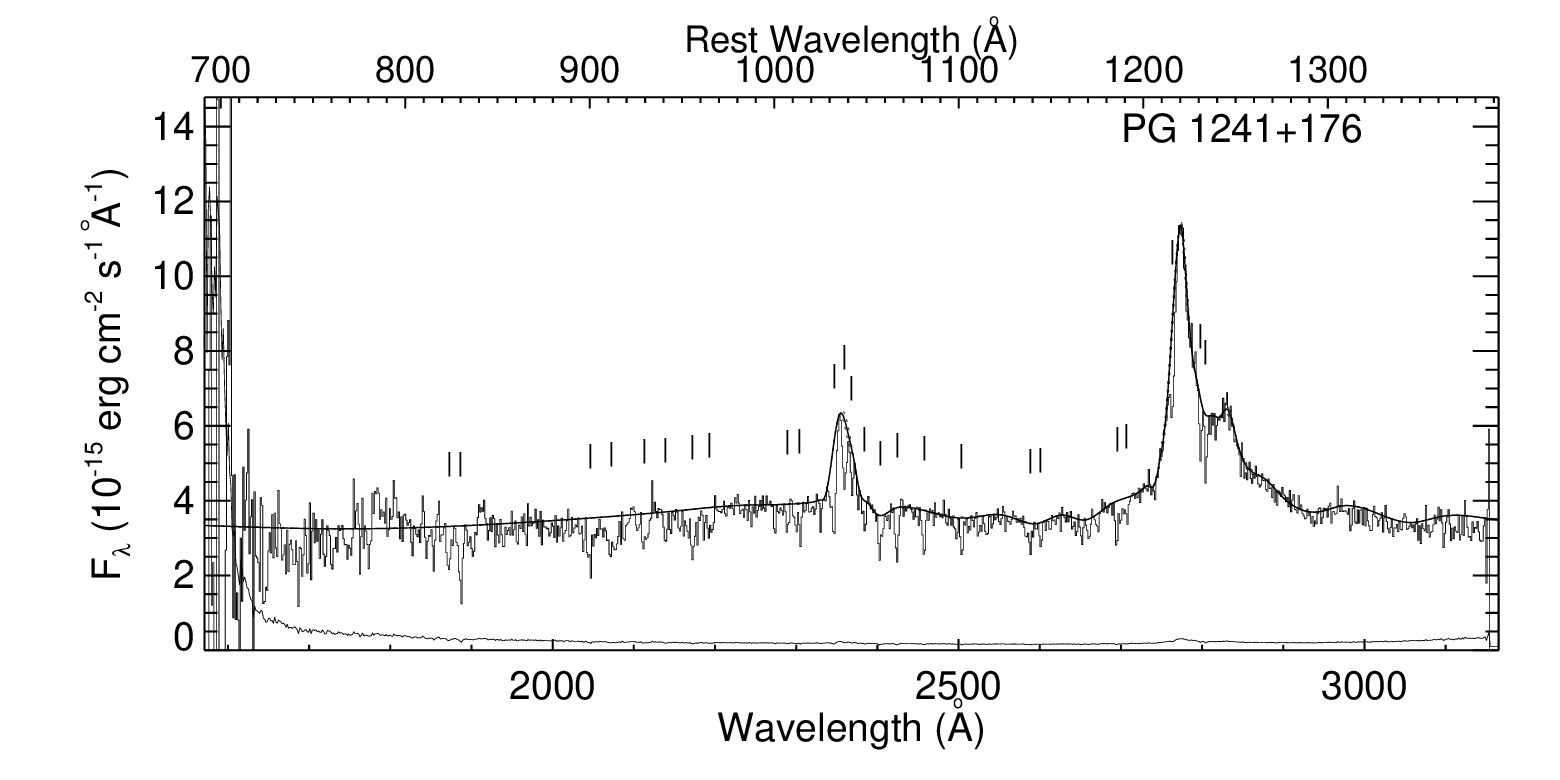}}
\centerline{\includegraphics[scale=0.4,angle=90]{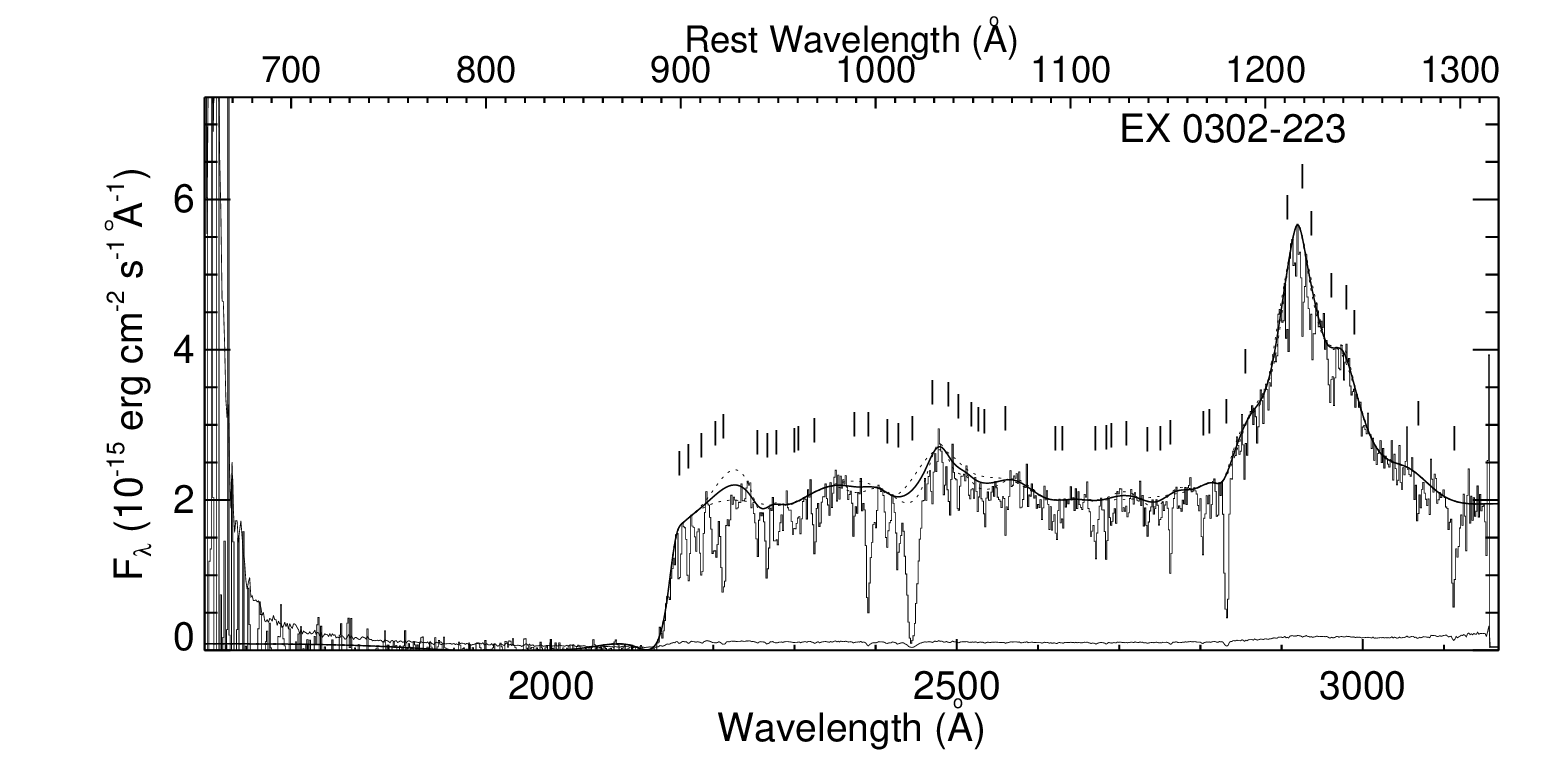}
            \includegraphics[scale=0.4,angle=90]{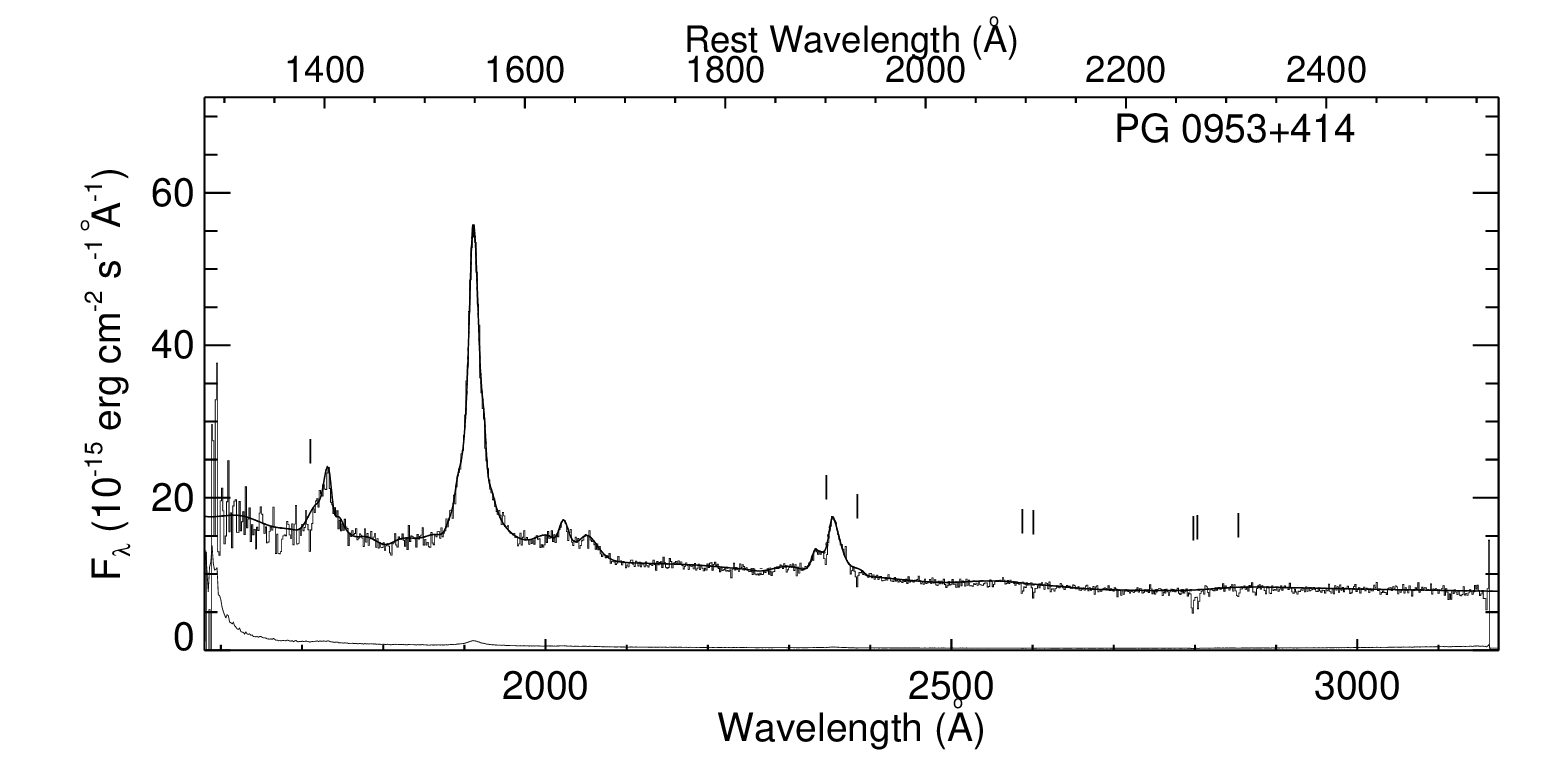}
            \includegraphics[scale=0.4,angle=90]{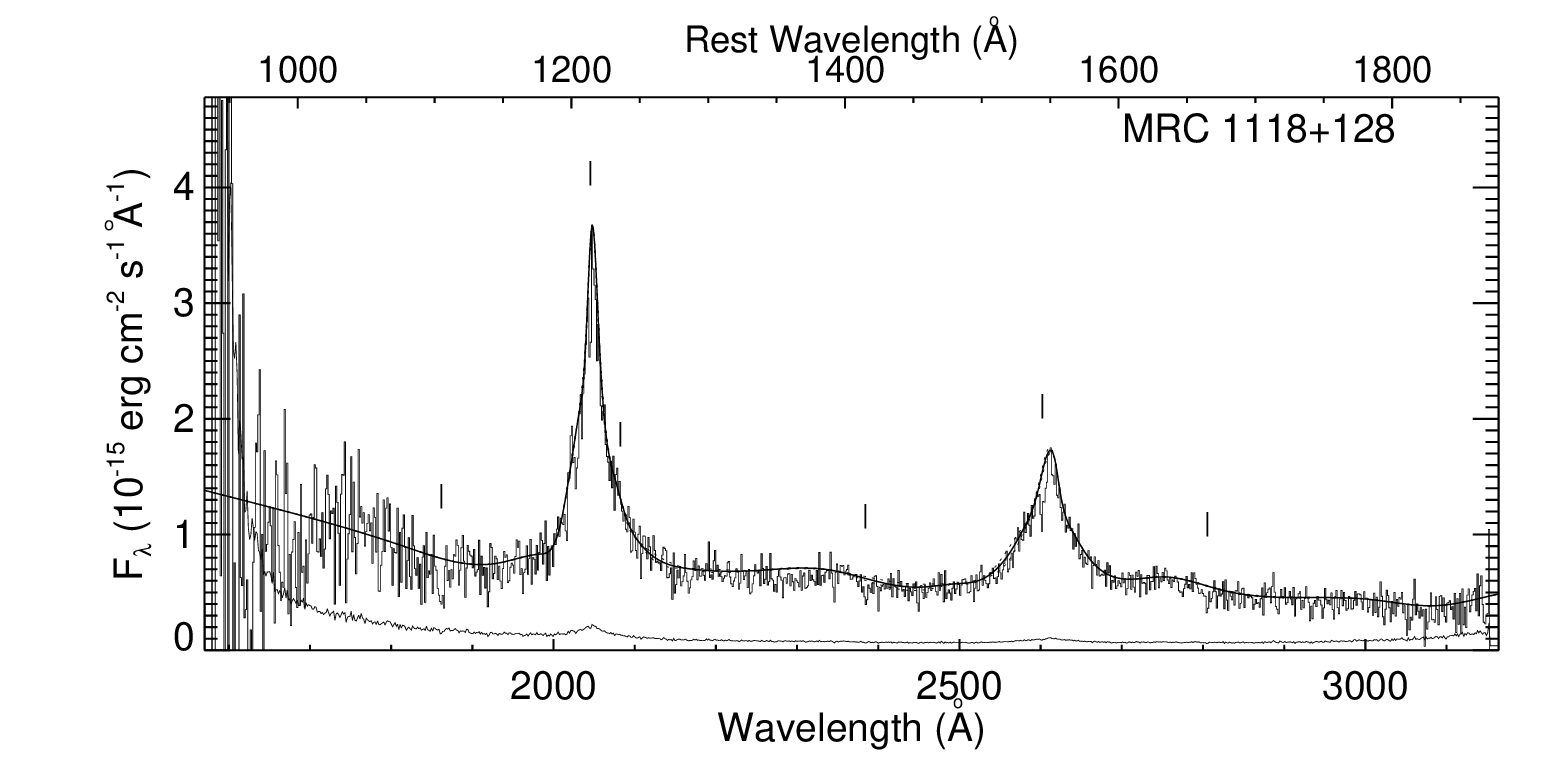}}
\caption{\small The STIS spectra of quasars observed in this survey with the
  best {\it effective} continuum fit superposed (the fit is not
  optimized in low-$S/N$ regions near the ends of some of the
  spectra).  The extreme continuum fits (see \S\ref{S_obs} of the
  text) are also shown as dotted lines.  The lower trace in each panel
  shows the error bar spectrum. The tick marks show the detected
  absorption lines listed in Table~\ref{tab_var} \label{fig_spec}}
\end{figure}

To reduce the STIS data, we used the {\sc calstis} pipeline
\citep{Brown02} in the Space Telescope Science Data Analysis System,
within the Image Reduction and Analysis Facility
(IRAF/STSDAS)\footnote{IRAF is distributed by the National Optical
Astronomy Observatories, which are operated by the Association of
Universities for Research in Astronomy, Inc., under cooperative
agreement with the National Science Foundation.}.  We did not
attempt the correct the zero-point of the STIS wavelength scale,
although we did determine its shift by comparing the wavelengths of
Galactic lines with those of their counterparts in FOS spectra; we
found that the STIS spectra were shifted by an average of 1.5~\AA\
(with a dispersion of 0.5~\AA).  Normalized spectra were created by
dividing the calibrated spectra with an {\it effective} continuum,
which includes both the true continuum and the broad emission lines.
The continuum fits were created by selecting points along the
spectrum, and fitting a cubic spline to those points.  We manually
selected the points so that the fit would effectively interpolate
across the NALs while conforming to the spectrum unaffected by the
NALs. We made no effort to optimize the continuum fit at the ends of
the spectra where the signal-to-noise ratio (hereafter $S/N$) was
low. In some cases, the absorption lines are broad and/or strong
enough that they irrecoverably distort the profiles of the emission
lines on which they are superposed. In such cases the placement of the
continuum can constitute a significant source of uncertainty in the
measured EW.  To include this uncertainty in our analysis, we used
two extreme continuum fits, to the FOS and to the STIS spectra, in
addition to the best fit.  The line identification procedure and EW
measurement were repeated for each of the fits.  An example of two
extreme continuum fits to the \ion{C}{4} region of the STIS spectrum
of PG~1309+355 is shown in Figure~\ref{fig_cont}. In
Figure~\ref{fig_spec} we show the STIS spectra of all objects with the
best continuum fit superposed as a solid line. The extreme continuum
fits are also shown as dotted lines, but these are not readily
discernible unless the differ considerably from the best fit.

After normalizing the spectra by dividing by the continuum fit, a
two-step process was carried out to detect and measure absorption
lines.  First, the unresolved line method \citep{Sc93}, used in the KP
project, was employed to detect lines in the spectra at the $5\sigma$
level \citep[i.e., at 5 ``significance levels'' in the formalism of
Schneider et al. 1993 and ][]{Jan98}. On a second pass, the EWs of the
identified lines were measured from the {\it observed} spectra by
integrating the data directly. The FOS and STIS spectra were searched
independently and the results were compared in the end. We measured
all lines in the spectral regions of interest, irrespective of their
origin.  Many of the targeted doublets that were resolved in the FOS
spectra were not resolved in the STIS spectra, therefore we cataloged
the two members of the doublet separately in the former case and
compared the sum of their EWs with the EW of the corresponding blends
in the STIS spectra. Similarly, in cases where lines of interest in
the STIS spectra were blended with lines from unrelated systems, we
did not attempt to de-blend them but measured the total EWs of the
blends and compared them with the sums of EWs of resolved lines in the
FOS spectra in the end.

A significant number of lines detected in the FOS spectra were not
detected in the STIS spectra, which typically had a lower $S/N$ than the
FOS spectra. In such cases, we determined and cataloged the local
photon noise level and the uncertainty resulting from the placement of
the continuum, which can be combined in quadrature to yield an upper
limit on the EW of the line in the STIS spectrum.

The results of the measurements are listed in Table~\ref{tab_var},
where we provide the following information:

\begin{description}

\item[\it Columns 1--4:] The observed wavelength and observed EW along
with the uncertainty in the EW, for lines measured in the STIS
spectra. The uncertainty is broken up into two parts, which should
be added in quadrature to give the final error bar on the EW: an
uncertainty due to Poison noise, $\sigma_{ph}$, and an uncertainty
due to a range of possible continuum fits, $\sigma_{cont}$
(discussed above and illustrated in Figure~\ref{fig_cont}). In cases
where the continuum fit was unambiguous, producing a negligible
uncertainty in the EW, we set $\sigma_{cont}=0$. If a line was
detected in a FOS spectrum but not in the corresponding STIS
spectrum, the EW column gives an upper limit, namely
$W_{\lambda}^{\rm max}({\rm STIS})=5\sigma_{ph}({\rm STIS})$.

\item[\it Column 5:] A set of flags indicating how blended lines in
the STIS spectra are compared with the corresponding resolved lines
in the FOS spectra. A ``$\Sigma$'' indicates that the line listed in
the STIS column is a known blend and the EW given is the total EW
of the blend. In the same row, under the FOS column, we give the sum
of the EWs of the individual lines making up the blend, as
measured in the FOS spectrum. The rows below the total EW row in
the FOS column give the EWs of the individual lines. These lines
are identified with increasing index numbers in column 5. The
corresponding rows in the STIS column are left blank.

\item[\it Columns 6--9:] The same information as in columns 1--4, but
for lines measured in the FOS spectra. We note that the there are only
two lines detected in STIS spectra but not in FOS spectra; for these 
lines we provide upper limits, analogous to the STIS upper limits.

\item[\it Columns 10 and 11:] The difference in equivalent width
between the FOS and STIS spectra: $\Delta W_{\lambda}\equiv
W_{\lambda}({\rm FOS}) - W_{\lambda}({\rm STIS})$, and this
difference normalized by its error bar, $\Delta
W_{\lambda}/\sigma$. The error bar was computed by adding all the
relevant error bars in quadrature, i.e., $\sigma^2 =
\sigma^2_{ph}({\rm STIS}) + \sigma^2_{cont}({\rm STIS}) +
\sigma^2_{ph}({\rm FOS}) + \sigma^2_{cont}({\rm FOS})$. In cases
where a line is detected in the FOS spectrum and not in the STIS
spectrum, we adopt $W_{\lambda}({\rm STIS})=W^{\rm
max}_{\lambda}({\rm STIS})\equiv 5\sigma_{phot}({\rm STIS})$ and
{\it vice versa} for a single case of a line detected with STIS but
not with FOS. Entries corresponding to variable absorption lines are
underlined in column 11 for easy identification.

\item[\it Columns 12--15:] The line identification, consisting of the
ion, the rest-wavelength of the transition, the difference between
the observed and expected wavelength of the line in the observer's
frame, $\Delta\lambda$, and the redshift of the absorption-line
system, $z_{a}$. The redshift of the system is determined from its
Ly$\alpha$ line, or from its strongest line, if Ly$\alpha$ is not
available.  The wavelength differences of other lines are determined
as $\Delta\lambda=(1+z_{a})\lambda_{rest}-\lambda_{obs}({\rm
FOS})$. In the case of blended doublets, we give identifications
for the constituent lines, rather than for the doublet. Entries
corresponding to variable absorption lines are underlined in column
12 for easy identification.

\item[\it Column 16:] The velocity offset of an absorption line for
cases where this is less than 10,000\kms.  A positive velocity
offset indicates a redshift of the absorption line relative to the
quasar. These velocities are based on wavelengths measured from the
FOS spectra whenever possible. Their uncertainties can be estimated
by considering the following: (a) typical uncertainties in the
absolute wavelength scales of the FOS spectra, which amount to
approximately 15\kms, (b) the fact that an uncertainty in the
redshift of order $\delta z\sim 1\times 10^{-4}$ results in an
uncertainty in velocity of no more than 30\kms, and (c) the
possibility that an additional heliocentric uncertainty of up to
30\kms\ may be needed. In summary, in cases where redshift
uncertainties are $\delta z \lsim 10^{-4}$, we estimate that
uncertainties in velocity offsets are of order 20--30\kms, while in
cases where redshifts uncertainties are $\delta z \sim 10^{-3}$, we
estimate that uncertainties in velocity offsets are of order
100--200\kms. Entries corresponding to variable absorption lines are
underlined for easy identification.

\end{description}

To check our methodology, we compared the our EW measurements from the
FOS spectra with those measured by \citet{Be02} from the same data and
found the two sets to be in very good agreement. This comparison was
facilitated by the fact that the results of \citet{Be02} are available
in a convenient electronic form\footnote{See {\tt
http://lithops.as.arizona.edu/$\sim$jill/QuasarSpectra/}.}. We
show the results of this comparison graphically in
Figure~\ref{fig_comp}, where we plot (a) the EW measured here against
the EW measured by \citet{Be02} and (b) the distribution of {\it
normalized} differences between our own measurements and those of
\citet{Be02}, namely $\Delta W_{\lambda}/\sigma \equiv
\left[W_{\lambda}({\rm theirs}) - W_{\lambda}({\rm ours}) \right]/
(\sigma^2_{\rm theirs} + \sigma^2_{\rm ours})^{1/2}$. This histogram
comprises 247 absorption lines, it appears symmetric about zero and
contains five outliers at $|\Delta W_{\lambda}/\sigma|>3$ (one is out
of the range of the plot). After examining closely the outlying cases
we can attribute all of them to differences in fitting the continuum.

\section{Results\label{S_res}} 

In the FOS spectra of the 15 quasars in our sample we have detected
and identified the following absorption lines: 62 associated
absorption lines (9 of which are slightly {\it redshifted} relative to
the quasar redshift), 56 non-associated lines, and 102 Galactic lines
(doublet members are counted separately in this census).  If we assign
the identified lines to {\it systems} according to their velocities,
we obtain 19 associated and 12 non-associated systems (4 of the
associated systems are redshifted relative to the quasar).  In
addition, we detected 67 unidentified lines in FOS spectra, the vast
majority of which are found at wavelengths blue-ward of the Ly$\alpha$
emission line in the spectra of the four highest-redshift quasars in
our sample. Thus, we attribute these lines to the Ly$\alpha$ forest
and related metal-line systems.

Closely spaced lines that are resolved in FOS spectra, are sometimes
blended into unresolved complexes in the STIS spectra because of the
lower spectral resolution and sampling rate of the latter. Moreover,
since the $S/N$ of the STIS spectra is lower than that of the FOS
spectra, many lines detected in the latter spectra are not detected in
the former. These two effects led to us finding 39 associated and 121
non-associated lines or complexes with FOS counterparts in the STIS
spectra (the complexes are identified by the flags in column 5 of
Table~\ref{tab_var}). We also find one associate and one
non-associated line in STIS spectra with only upper limits from FOS
spectra.  Moreover, for 86 lines/complexes in the FOS spectra we were
only able to obtain EW upper limits from the STIS spectra. Finally, we
also detected 21 lines in regions of the STIS spectra not covered in
the FOS spectra.

To detect NAL variability, we compared the EW difference between
observations with the uncertainty resulting from photon noise and from
the placement of the continuum. In particular, we computed the
normalized EW difference, $\Delta W_{\lambda}/\sigma$, as defined in
the previous section and tabulated it in column 11 of
Table~\ref{tab_var}, for all lines or blended complexes detected in
both the STIS and FOS spectra (in the case of blended complexes we
formed the difference between the EW of the complex from STIS and the
sum of EWs of constituent lines from FOS).  We show the results of
this comparison graphically in Figure~\ref{fig_scatter}, where we plot
(a) the STIS and FOS EWs against each other, and (b) the distribution
of the normalized EW difference. We divide the lines into two
categories, for which we make separate plots: associated lines,
defined by $|\Delta v|<5,000$\kms \citep[relative to the quasar
redshift; following][]{Fo86}, and non-associated lines (including
Galactic and unidentified lines). Included in this figure are the two
lines with STIS detections and FOS upper limits, bringing the total
number of lines to 40 associated and 122 non-associated.  It is
noteworthy that out of 122 non-associated lines the difference in EW
between the FOS and STIS spectra never exceeds $3\sigma$.

The distribution of normalized EW differences in the upper panel of
Figure~\ref{fig_scatter}b shows a number of outliers at $|\Delta
W_{\lambda}|/\sigma > 3$, which we interpret as variable lines.  We
find 3 such outliers, when the expectation value, assuming a Gaussian
probability distribution, is 0.11.  These are the \lion{C}{4}{1549}
doublets in PKS~2135--14, MRC~2251--178, and PG~2251+113 ($\Delta
W_{\lambda}/\sigma =3.4$, 3.6, and 5.0, respectively).  Absorption
lines that are deemed to be variable are identified in
Table~\ref{tab_var} by underlining the value of $\Delta
W_{\lambda}/\sigma$ as well as the line identification and velocity
offset.

To search for additional variable lines, we compared the 5$\sigma$ EW
upper limits of the 86 lines not detected in the STIS spectra with the
EWs of their counterparts measured from the FOS spectra. The results
of this comparison are illustrated in Figure~\ref{fig_limits}, where
we plot the distribution of $[W_{\lambda}({\rm FOS})-W^{\rm
    max}_{\lambda}({\rm STIS})]/(\sigma^2_{\rm FOS} +\sigma^2_{\rm
  STIS})^{1/2}$. This is analogous to the distribution of normalized
EW differences plotted in Figure~\ref{fig_scatter}, but with
$W_{\lambda}({\rm STIS})$ replaced by the upper limit $W^{\rm
  max}_{\lambda}({\rm STIS})\equiv5\sigma_{ph}({\rm STIS})$. The plot
shows one clear outlier with a 3.3$\sigma$ deviation, which is the
Ly$\alpha$ line in PG~1718+481 at $\Delta v=+469$\kms, which we take
as a variable line.

We have scrutinized the EW measurements of apparently variable NALs to
make sure that their variability is not caused by systematic errors
such as continuum placement or line blends in STIS spectra. We also
compared our FOS EW values of the apparently variable lines with those
of \citet{Be02} and found them to agree to ``$1\sigma$'' or better in
the cases of PG~1718+481, MRC~2251--178, and PG~2251+113.  As a case
in point, we note that the variability of the \ion{C}{4} doublet of
MRC~2251--178 has already been reported and discussed by
\citet*{Ga01}. The analysis of this paper was carried out
independently of the that paper and yielded a similar result.  In the
case of PKS~2135--14, our measurement differs from that of
\citet{Be02}, but we believe that our own measurement is the one that
should be adopted. We discuss this case in detail and present our
arguments in \S\ref{S_indiv}.  To provide a visual demonstration of
variable lines we overlayed the normalized spectra from the FOS and
STIS. Before carrying out this exercise, we convolved the FOS spectra
with a Gaussian of width equal to one STIS resolution element.  To
account for the sampling differences between the two instruments, we
discarded every other pixel in the FOS spectra to reduce the sampling
rate to two pixels per resolution element.  We then interpolated the
FOS spectra using a cubic spline so that the final wavelength bins
matched those of the corresponding STIS spectra. Finally, we aligned
the wavelength scales of corresponding spectra by cross-correlating
regions containing Galactic absorption lines. The results of this
exercise are shown in Figure~\ref{fig_overp}, where we plot segments
of the original STIS and FOS spectra as well as the normalized spectra
from the two instruments (after smoothing, resampling, and alignment).

\section{Discussion and Conclusions}

\subsection{Summary of Results and Immediate Implications\label{S_summ}}

We can summarize our main observational results as follows: 

\begin{enumerate}

\item
We have identified four variable associated absorption lines. None of
the non-associated lines were found to vary. The EWs of the variable
lines span the entire range of EWs in the absorption-line sample.

\item
Each of these lines traces a different absorption {\it system},
implying that at least 21\% (4/19) of associated systems are
intrinsic.

\item
The variable lines are very close to the quasar emission redshift:
four out of the five have $|\Delta v|<500$\kms. Moreover, one of the
variable systems is redshifted relative to its host quasar.

\end{enumerate}

Since not all intrinsic lines will have necessarily varied between
observations, the fraction of variable systems represents a lower
limit on the fraction that originate in gas in the immediate vicinity
of the quasar central engine.  Our result is comparable to that of
\citet{Na04} who find that at 23\% of the associated NALs systems that
they have monitored from the ground in $z\sim 2$ quasars are variable.
Moreover, \citet{Ba97} report a similar fraction of variable NALs
($\sim 30$\%) in higher-redshift ($z\gsim 2.5$) quasars.  This
estimate of the fraction of intrinsic NAL systems is also compatible
with the estimate of 50\% for $z_{\rm e} \approx 2$ quasars by
\citet{Ga99}, who used partial coverage as their diagnostic tool.  We
do note the caveat that the studies of \citet{Ga99} and \citet{Na04}
were based on a small number of objects (6 in the former and 8 in the
latter), therefore their exact statistical results are somewhat
uncertain.

The variability we observe could result from one of the following two
processes.

\begin{enumerate}

\item{\em Fluctuations of the Quasar Light. --} Fluctuations of the
  quasar's ionizing continuum induce the ionization structure of the
  gas to vary.  One can test this hypothesis by studying the
  variability of lines from ions with different ionization potentials,
  as we discuss further below.  Absorption features should have a
  delayed reaction to the variations of the quasar's ionizing
  radiation due to the finite recombination time of the absorbing gas.
  In our study and in the work of \citet{Ha95}, \citet*{Ald97}, and
  \citet{Na04}, NAL variability time scales are typically about 3--5
  years in the quasar rest frame.  If we take these to be upper limits
  to the recombination time, we can estimate a lower limit of the
  electron density of the absorber clouds by these time scales,
  following \citet{Ha97b}. The recombination time is $\tau_{\rm
    rec}\sim 1/\alpha_{\rm r} n_{\rm e}$, where $\alpha_{\rm r}$ is
  the recombination rate coefficient and $n_{\rm e}$ is the electron
  density.  For the \ion{C}{4} ion, $\alpha_{\rm r} = 2.8 \times
  10^{-12}~{\rm cm^{-3}}~{\rm s^{-1}}$ \citep{Ar85}, which leads to
  $n_{\rm e}\gsim 3,000~{\rm cm^{-3}}$.  Using this density
  constraint, we can place limits on the distance of the absorber from
  the ionizing source. We adopt equation (3) of \citet{Na04} for the
  distance of the absorber from the ionizing source, which we recast
  as $R\approx 95\; L_{44}^{1/2}\; n_4^{-1/2}\; U_{-2}^{-1/2}$~pc,
  where $U_{-2}$ is the ionization parameter in units of $10^{-2}$,
  $L_{44}$ is the bolometric luminosity of the quasar in units of
  $10^{44}$~erg~s$^{-1}$, and $n_4$ is the density in units of
  $10^4~{\rm cm^{-3}}$. Following \citet{Na04}, we estimate the
  bolometric luminosity as $L_{\rm bol}\approx 4.4\lambda L_{\lambda}$
  at $\lambda=1450$~\AA; we obtain $L_{44}\approx 1$ for the quasars
  with variable \ion{C}{4} lines. Finally, if we assume the optimal
  ionization parameter, $U\approx 0.02$, from the models of
  \citet{Ha95} and use the above density constraint, we obtain $R\lsim
  100$~pc.

\item{\em Bulk or Internal Motion of the Absorber. --} Bulk motion of
  the absorber across the quasar line of sight will cause the
  absorption features to vary for one or both of the following
  reasons: (a) a change in the column density of the absorber, and/or
  (b) a change in the coverage fraction in the case of a patchy
  absorber that covers the background source only
  partially. Similarly, density changes along the line of sight to the
  continuum source (resulting form the passage of pressure waves, for
  example) would have the same effect for the same reasons.  The
  observed variability time scales could thus be equally well
  interpreted in this context. However, for a unique interpretation,
  we need more information about the location and physical conditions
  of the absorber. As a specific example, we consider a scenario in
  which the absorber is a small parcel of gas in the broad
  emission-line region (BELR). Assuming the continuum source is the
  UV-emitting region of the inner accretion disk, its size is of order
  $D_{\rm cont}\sim 10\; GM/c^2 = 1.5\times 10^{14}\; M_8$~cm, where
  $M_8$ is the mass of the black hole in units of $10^8~{\rm
    M}_{\odot}$. If a parcel of gas crosses the cylinder of sight to
  this region at the dynamical speed of the BELR, $v_{\rm dyn}\sim
  (GM/r)^{1/2}\sim 3,600\; M_8^{1/2}\; r_{17}^{-1/2}$\kms\ \citep[where
    $r_{17}$ is the radius of the BELR in units of $10^{17}$~cm;
    see][]{Kaspi00}, the variability time scale will be of order
  5~days. If, on the other hand, a pressure wave crosses the cylinder
  of sight at the speed of sound ($c_{\rm s} \approx 10\;
  T_4^{1/2}$\kms, with $T_4$ the temperature in units of $10^4$~K),
  the variability time scale will be of order 5~years. If we assume
  that the absorber is further away from the quasar central engine
  (e.g., in the narrow-emission line region or in the host galaxy)
  the above estimates can be modified accordingly.

\end{enumerate}

The moral of the above discussion is that the origin of
absorption-line variability can be diagnosed with the right
observations. In particular, if the variability time scales are
constrained to be on the order of a few months or less, internal
motions (i.e., waves) in the absorber will be called into question.
Such stringent constraints have, in fact, been obtained for some
high-redshift quasars \citep[see][and references therein]{Na04}.
Another test can be performed using high-dispersion spectra covering
transition from a wide range of ionization states, as we discuss in
detail in \S\ref{S_fut}, below.

\subsection{Discussion of Individual Objects\label{S_indiv}}

\begin{description}

\item{\em EX 0302--223. --} The redshift quoted for this object in
  quasar catalogs is 1.400. This value was taken from \citet*{Cha81}
  and is incorrect. The correct redshift is 1.409 and is reported by
  \citet*{Marg85}, who also note that the former value had a
  typographical error. The difference in the two redshifts translates
  into a velocity difference of about 1100\kms, which would have led
  to a significant error in the velocities of Table~\ref{tab_var}. In
  fact, the absorption system at $\Delta v \approx -450$\kms\ would
  have appeared at a positive velocity if the former value of the
  redshift were used.

\item{\em QSO J0909--095. --} Although the \ion{C}{4} doublet is
  detected in the FOS spectrum, the poor $S/N$ and lower sampling of
  the STIS spectrum lead to a large and uninteresting upper limit to
  the EW at the second epoch. We do, however, detect the Ly$\alpha$
  line in the STIS spectrum. The redshift of this object comes from a
  low-resolution spectrum by \citet{Kne98} in which no narrow,
  forbidden lines were detected. Therefore, it is unusually uncertain
  and is quoted only to two decimal places.

\item{\em QSO 0957+561A. --} A damped Ly$\alpha$ absorber (DLA) at
  $z=1.391$ along this line of sight produces many NALs \citep[see,
  for example,][]{Dol00}.  As a result, continuum fitting is uncertain
  throughout the spectrum, which restricts our conclusions about the
  nature of the associated absorber. This object is the prototypical
  gravitationally-lensed quasar, discovered by \citet*{Wal79}. It is a
  radio-loud quasar according to its 5~GHz power \citep[$P_{\rm 5\;
  GHz} =6\times 10^{25}~{\rm W~Hz^{-1}}$ based on the flux reported
  by][]{Has81}. The redshift reported in quasar catalogs is 1.4136,
  which is the value measured from the broad \ion{Mg}{2} emission line
  by \citet{We79b}. We believe that this value does not correctly
  reflect the redshift of the quasar for the following reasons. There
  are several measurements of the redshift based on the broad
  \ion{C}{4}, \ion{C}{3}] and \ion{Mg}{2} UV emission lines
  \citep{Wal79,We79b,Wil80,You81}. Measurements from the same emission
  line by different authors agree with each other but redshifts from
  the \ion{C}{4} and \ion{C}{3}] lines are systematically lower than
  the redshift from the \ion{Mg}{2} line by about $3\times
  10^{-3}$. This is consistent with a well known trend in radio-loud
  quasars, in which the centroid of the broad \ion{C}{4} line is at
  the same redshift as the narrow, forbidden lines, while the the
  centroids of the broad \ion{Mg}{2} and H$\beta$ lines are slightly
  redshifted \citep[see, for example,][]{Bro94,Mar96}. Therefore, we
  believe that the redshift obtained from the \ion{C}{4} emission line
  is more likely to represent the systemic redshift of the quasar
  and we adopt the value of 1.4093
  from \cite{You81}. If the former redshift were adopted, the
  velocities listed in the last column of Table~\ref{tab_var} would
  increase by 510\kms.

\item{\em PKS 2135--14. --} The \ion{C}{4} absorption line of this
  object has been the subject of detailed study by several authors.
  It is an interesting case because the absorption lines are {\it
  redshifted} relative to the peaks of the broad emission lines and
  relative to the systemic redshift, as defined by the narrow,
  forbidden lines.  In particular, \citet{BK83} compared {\it IUE}
  spectra taken approximately 2.6 years apart and found a 25\% change
  in EW (from 2.0~\AA\ in 1979 to 2.5~\AA\ in 1981) as well as a shift
  of the absorption-line centroid by 300~\kms\ between the two
  epochs. We find a significant change in EW between the STIS and FOS
  observations (from 4.7~\AA\ in 1992 to 3.2~\AA\ in 2000; see
  illustration in Fig.~\ref{fig_overp}), as well as a significant
  change between the {\it HST} (1992 and 2000) and {\it IUE} (1979 and
  1981) observations.  However, the EW we measure from the FOS
  spectrum is higher than the values measured by \citet{Be02} and
  \citet{Ha97c} (4.7~\AA\ by us {\it vs} 3.1~\AA\ and 3.6~\AA\ by
  them). The difference is a result of the placement of the {\it
  effective} continuum (i.e., the emission line profile). We have
  experimented with several different fits to the peak and blue side
  of the \ion{C}{4} emission-line profile including using the profile
  of the \ion{C}{3}]$\;\lambda$1909 line as a template.  In
  Figure~\ref{fig_pksfits} we show the \ion{C}{4} profile of this
  object as observed with the FOS, with our two extreme continuum fits
  superposed as smooth, solid lines. The fit based on the \ion{C}{3}]
  profile falls between the two extremes and is the one we adopt as
  the optimal. We obtained the EW by integrating the normalized
  spectrum directly between the two extreme points where the fitted
  profile meets the observed profile.  In contrast, the fit used by
  \citet{Ha97c} resembles the shape of our lowest acceptable fit but is
  placed somewhat lower (see their Fig.~3); it suffers from the
  drawback that the resulting peak of the \ion{C}{4} line falls a few
  {\AA}ngstrom short-ward of its nominal wavelength. Those authors
  obtained the EW by fitting the absorption profile with 2 or 4
  components. The fit used by \citet{Be02} was a simple interpolation
  over the primary absorption trough (illustrated schematically as a
  dotted line in Fig.~\ref{fig_pksfits}), which also led to a low
  value for the EW. \citet{BB86} and \citet{Ha97c} have studied the
  ionization conditions in this particular absorption system and
  concluded that the ionization parameter is likely to be high
  ($U\approx 0.25$), although lower values (as low as $U\approx
  3\times 10^{-3}$) could not be ruled out. They discussed a number of
  possible scenarios for the absorber, including galaxies within the
  cluster harboring the quasar, the quasar host galaxy itself, and gas
  intrinsic to the quasar central engine, but they were not able to
  select one of these scenarios as an obvious favorite. The
  variability results we present here, especially when combined with
  the earlier results of \citet{BK83}, make a strong case that this
  absorber is intrinsic to the quasar central engine. This finding is
  particularly interesting in view of the fact that this
  absorption-line system appears to be redshifted relative to the
  quasar. Such a redshift does not necessarily imply infall of the
  absorbing gas toward the quasar central engine because other
  scenarios can be found that produce such a redshifted absorption
  line. For example, a parcel of gas in a rotating accretion-disk wind
  passing in front of an {\it extended} continuum source at the center
  of the disk can easily produce a redshifted absorption line.

\end{description}

\subsection{Future Prospects\label{S_fut}}

The variability of lines from ions of different ionization potentials
can be used to distinguish between variations in the ionizing
continuum and variations of the column density (due to transverse
motion of the absorber). Ideally, high-resolution ($R \gsim 40,000$)
spectra should be used for such a test so that the absorption lines
are fully resolved. Under these conditions one can determine whether
the background source is partially covered and infer the column
density of the absorber, thus constraining its location \citep[see,
for example][]{BaSa97,Ha97a,Ga99}.

To illustrate and evaluate the method, we carried out the following
exercise.  We used the photoionization code CLOUDY \citep{Fer96} to
simulate the the ionization structure of the absorber, assuming that
the incident ionizing spectrum is that of \citet{MaFe87}.  Input
parameters for the models are the total hydrogen density, the
ionization parameter, and the total hydrogen column density for each
component, assuming Solar abundances and a metallicity of 1/3 the
Solar value.
Throughout this exercise we assumed that the coverage fraction is
unity for all transitions and that it did not change as the column
density or ionization parameter changed.  The results for simulated
profiles are not sensitive to the hydrogen density in the optically
thin regime; values from $10^5$ to $10^9~{\rm cm}^{-3}$ give similar
results.  For our particular example we have assumed that the absorber
consists of three kinematic ``components'' at relative velocities
$-75$, 0, and $20\kms$, with column densities of $\log N_{\rm H}
=19.5$, 20, and 19, and broadening parameters $b = 8$, 12, and
10\kms. The ionization parameter is taken to be $\log U =-2$.  In
Figure~\ref{fig_synth} we display the synthetic spectra resulting from
this exercise. The middle set of panels show a simulated spectrum
($R=23,500;~S/N=20$) of the \ion{C}{4} profile along with a variety of
other transitions.  The left-hand set of panels of
Figure~\ref{fig_synth} demonstrate the effect of bulk motion, which we
represent as a decrease of the column density of each absorption
component by an order of magnitude.  The result is that all of the
absorption lines become weaker, regardless of ionization state.  The
right-hand set of panels demonstrates the effect of increasing the
continuum strength and hence the ionization parameter by an order of
magnitude. The lines from the lower ionization species are not
detected, while the lines from the higher ionization species, such as
\ion{N}{5} and \ion{O}{6} become stronger.  A decrease in the total
column density and an increase in the continuum strength have a
similar effect on the \ion{C}{4} profile, but the difference between
the two scenarios can be diagnosed if multiple transitions from a
variety of ionization states are observed.

With the above scientific questions and technical considerations in
mind, the next logical step in our systematic study of intrinsic NALs
is to target specific quasars whose NALs are demonstrably intrinsic.
If the quasars cooperate, we can infer the physical conditions and
location of the absorber through repeated observations. Thus we will
be able to assess the role of the NAL gas in the overall accretion
flow.  Alternatively, significant progress can also be made by single
epoch observations of intrinsic (and other, associated) NALs at high
spectral resolution. Using high-resolution spectra that encompass
transitions from a wide range of ionization states, we can determine
the ionization structure of the absorber and constrain its distance
from the ionizing source.

\acknowledgments 

This work was supported by grant HST-GO-08681.01-A from the Space
Telescope Science Institute, which is operated by the Association of
Universities for Research in Astronomy, Inc., under NASA contract
NAS5-26555. We also acknowledge support from NASA grant NAG5-10817.
We have made use of the NASA/IPAC Extragalactic Database (NED) which
is operated by the Jet Propulsion Laboratory, California Institute of
Technology, under contract with the NASA. We thank the anonymous
referee for many valuable technical comments. We are grateful to Sofia
Kirhakos and Buell Jannuzi for providing us with the uniformly and
fully reduced spectra of quasars observed by the {\it HST}/FOS. We
also thank Fred Hamann for useful discussions and especially for
reminding us of the importance of accurate redshifts.



\begin{deluxetable}{lllcccll}
\tabletypesize{\footnotesize}
\tablecolumns{8}
\tablewidth{0in}
\tablecaption{Journal of Observations\label{tab_obs}}
\tablehead{
\colhead{} & \colhead{} & \colhead{} & \multicolumn{2}{c}{Obs. Date (UT mm/dd/yyyy)} & \colhead{} 
& \colhead{$z$} & \colhead{Pub} \\
\noalign{\vskip -6pt}
\colhead{} & \colhead{} & \colhead{} & \multicolumn{2}{c}{\hrulefill} & \colhead{Interval} & \colhead{} \\
\colhead{Object} & \colhead{$z$\tablenotemark{a}} & 
\colhead{Targeted Lines} & \colhead{FOS} & \colhead{STIS} & \colhead{(years)} & 
\colhead{Refs\tablenotemark{b}} & \colhead{Refs\tablenotemark{c}} }
\startdata
EX 0302--223  & 1.409      & \ion{N}{5}, Ly$\alpha$, \ion{O}{6} & 12/07/1995 & 02/22/2001 & 5.22 & 1    & 1    \\ 
QSO J0909--095& 0.63       & \ion{C}{4}                         & 10/29/1996 & 03/04/2001 & 4.35 & 2    & none \\ 
PG 0953+414   & 0.2341(4)  & \ion{C}{4}        & 11/05/1991\tablenotemark{d} & 01/21/2001 & 9.22 & 3    & 2,3  \\ 
              &            &                   & 06/18/1991\tablenotemark{d} &            & 9.60 &      &      \\ 
QSO 0957+561A & 1.4093(1)  & \ion{N}{5}, Ly$\alpha$, \ion{O}{6} & 01/26/1995 & 09/08/2000 & 5.62 & 4    & 4    \\ 
MRC 1118+128  & 0.685(1)   & \ion{C}{4}, Ly$\alpha$             & 02/26/1995 & 03/28/2001 & 6.09 & 5    & none \\ 
PG 1241+176   & 1.273      & \ion{O}{6}, Ly$\alpha$             & 12/08/1992 & 03/19/2001 & 8.28 & 6    & 2    \\ 
PG 1309+355   & 0.184      & \ion{C}{4}                         & 05/20/1996 & 04/04/2001 & 4.88 & 6,7  & 5    \\
\enddata
\tablenotetext{a}{See \S\ref{S_obs} and \S\ref{S_indiv} of the text for a
discussion of the adopted emission redshift values and related velocity
uncertainties.  The figure in parethesis gives the uncertainty in the
last digit whenever this is available.}
\tablenotetext{b}{{\em References to adopted redshift measurements. --}
(1) \citet*{Marg85};
(2) \citet{Kne98};
(3) \citet{Mar96};
(4) \citet{You81};
(5) \citet{Wil76};
(6) \citet{Sch83};
(7) \citet{Bor92};
(8) \citet{Era04};
(9) \citet{Ha97c};
(10) \citet{BB86};
(11) \citet{Ber83}.
}
\tablenotetext{c}{{\em References to previously published FOS spectra}. -- 
(1) \citet{Boi98}; 
(2) \citet{Jan98}; 
(3) \citet{Wil99}; 
(4) \citet{Dol00}; 
(5) \citet*[][the spectra were discussed but not displayed]{Bra00}; 
(6) \citet{Mar96}; 
(7) \citet{Bah93}; 
(8) \citet{Ha97c}; 
(9) \citet{Ga01}; 
(10) \citet{Net95}.
}
\tablenotetext{d}{The earlier date corresponds to the FOS G190H spectrum and the later date the FOS G270H spectrum.}
\end{deluxetable}
\clearpage
\input tab2.tex

\clearpage
\clearpage
\clearpage
\begin{figure}  
\centerline{
            \includegraphics[scale=0.4,angle=90]{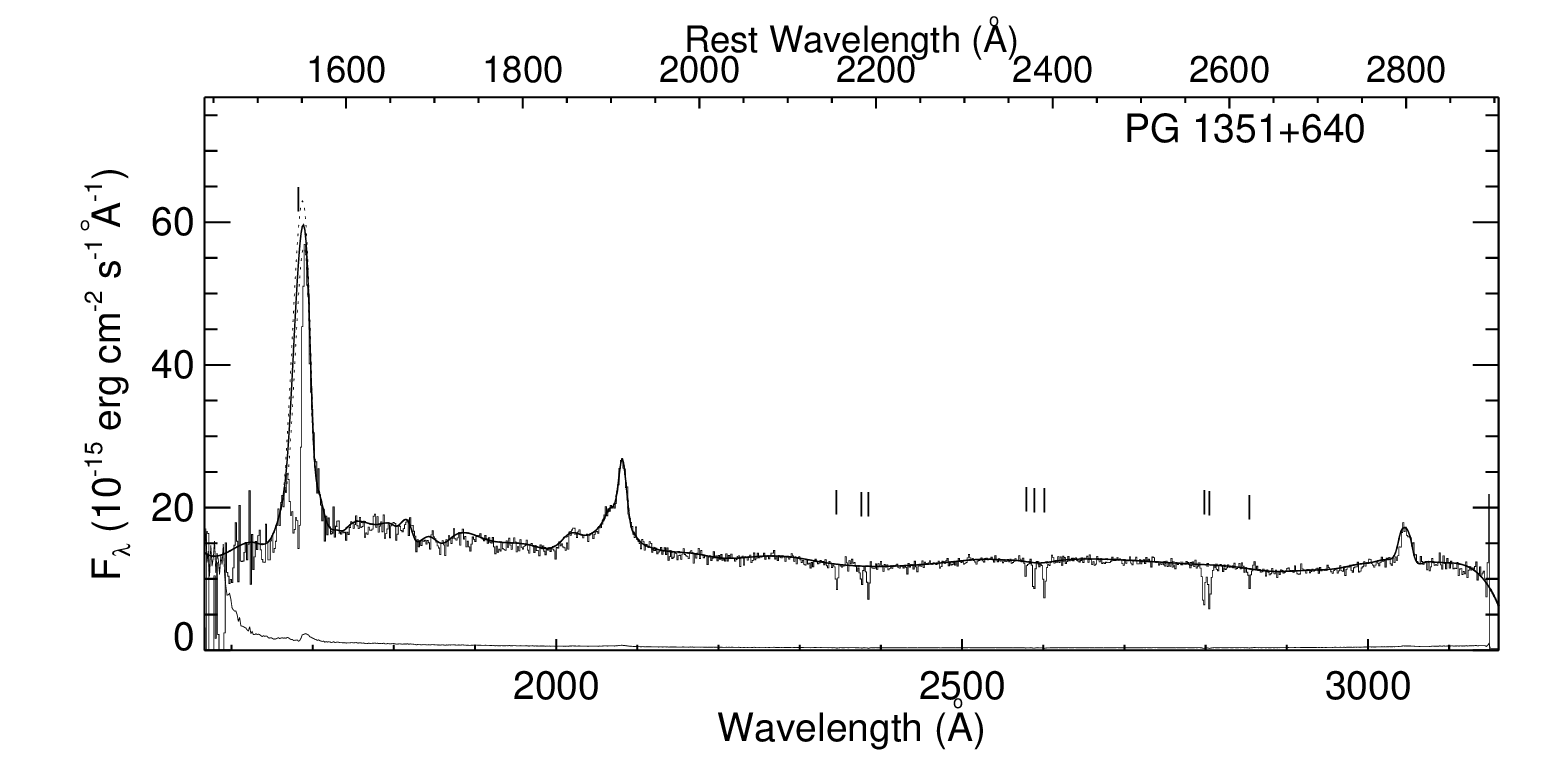}
            \includegraphics[scale=0.4,angle=90]{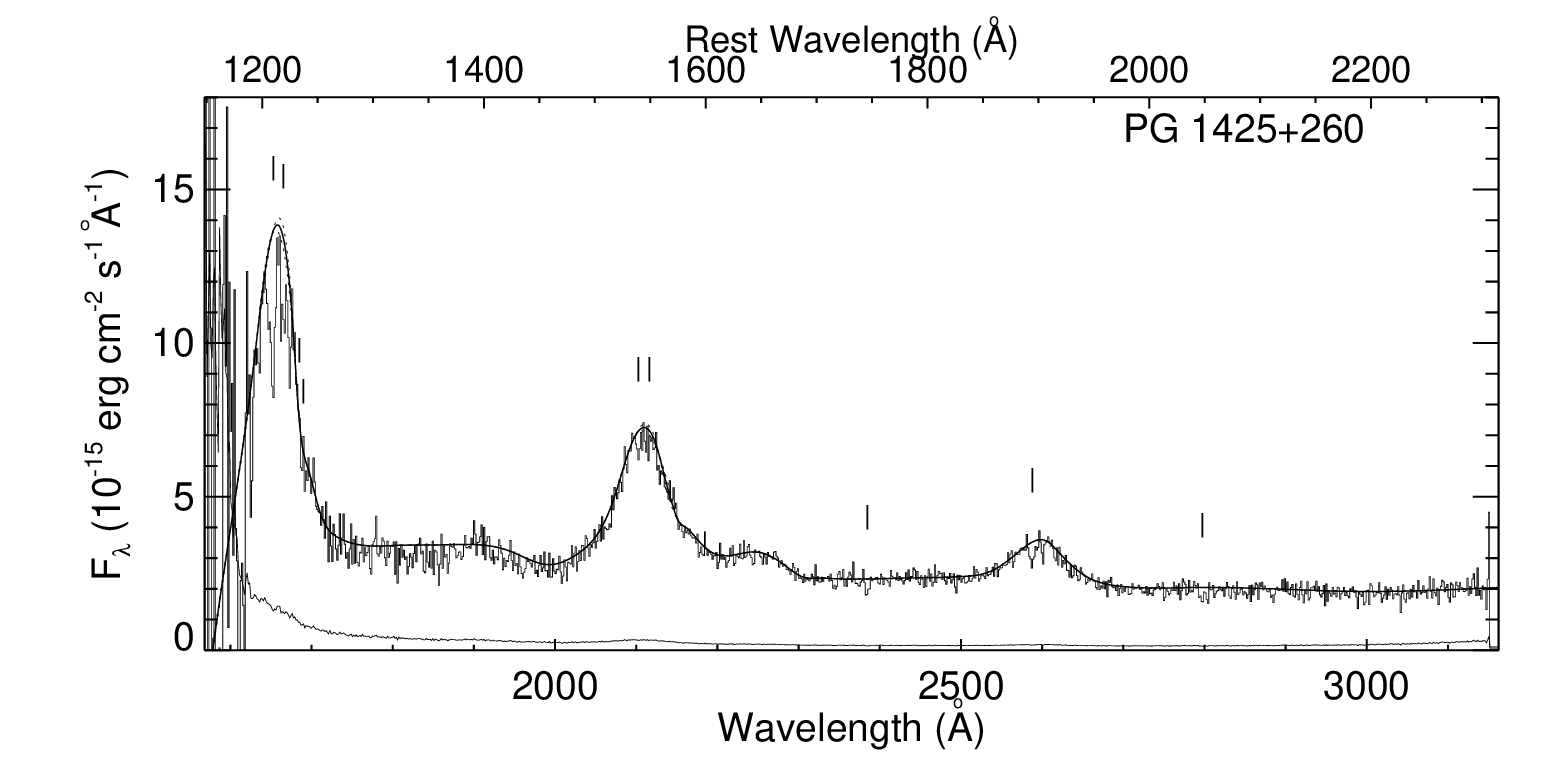}
            \includegraphics[scale=0.4,angle=90]{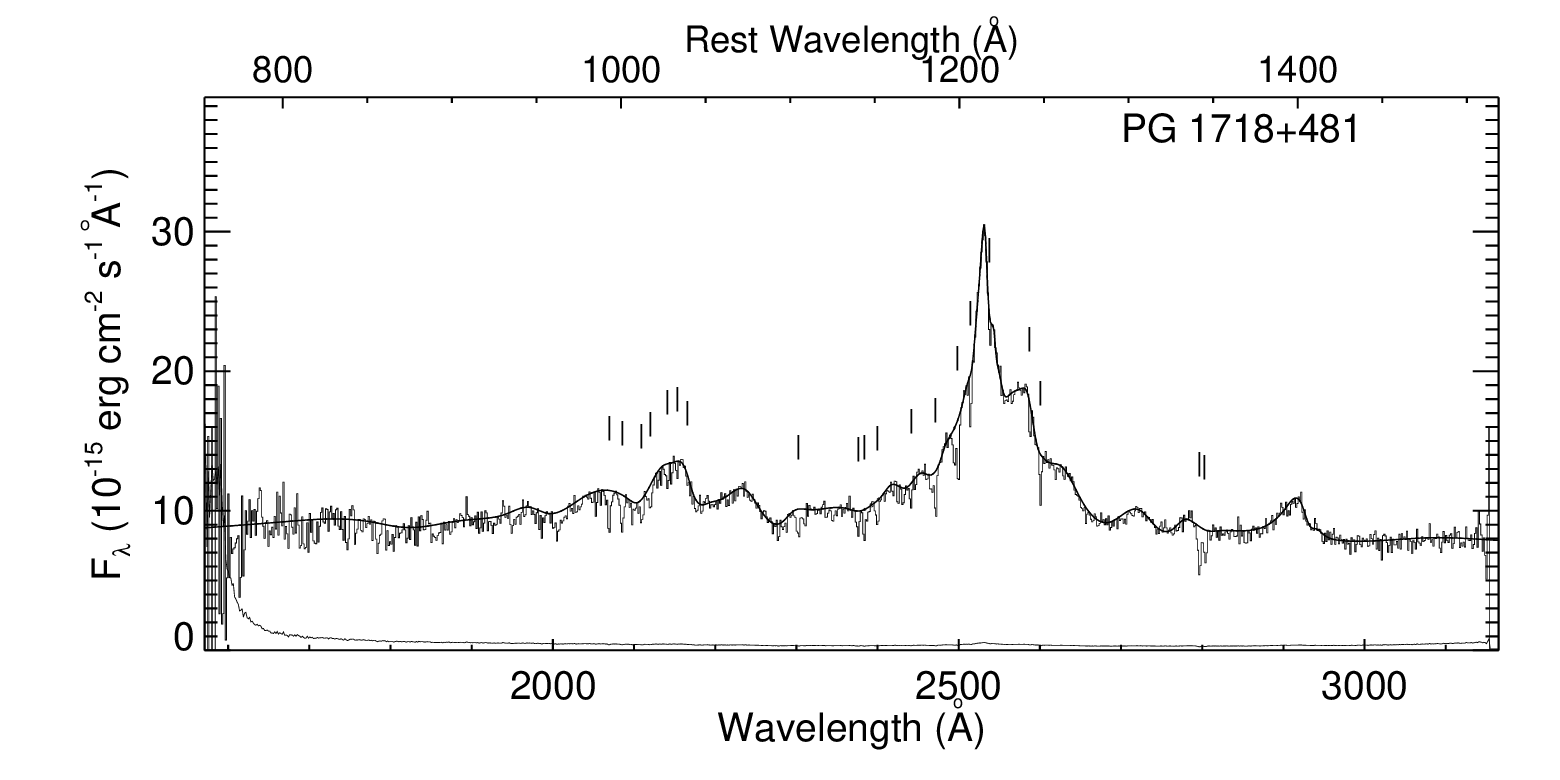}}
\centerline{
            \includegraphics[scale=0.4,angle=90]{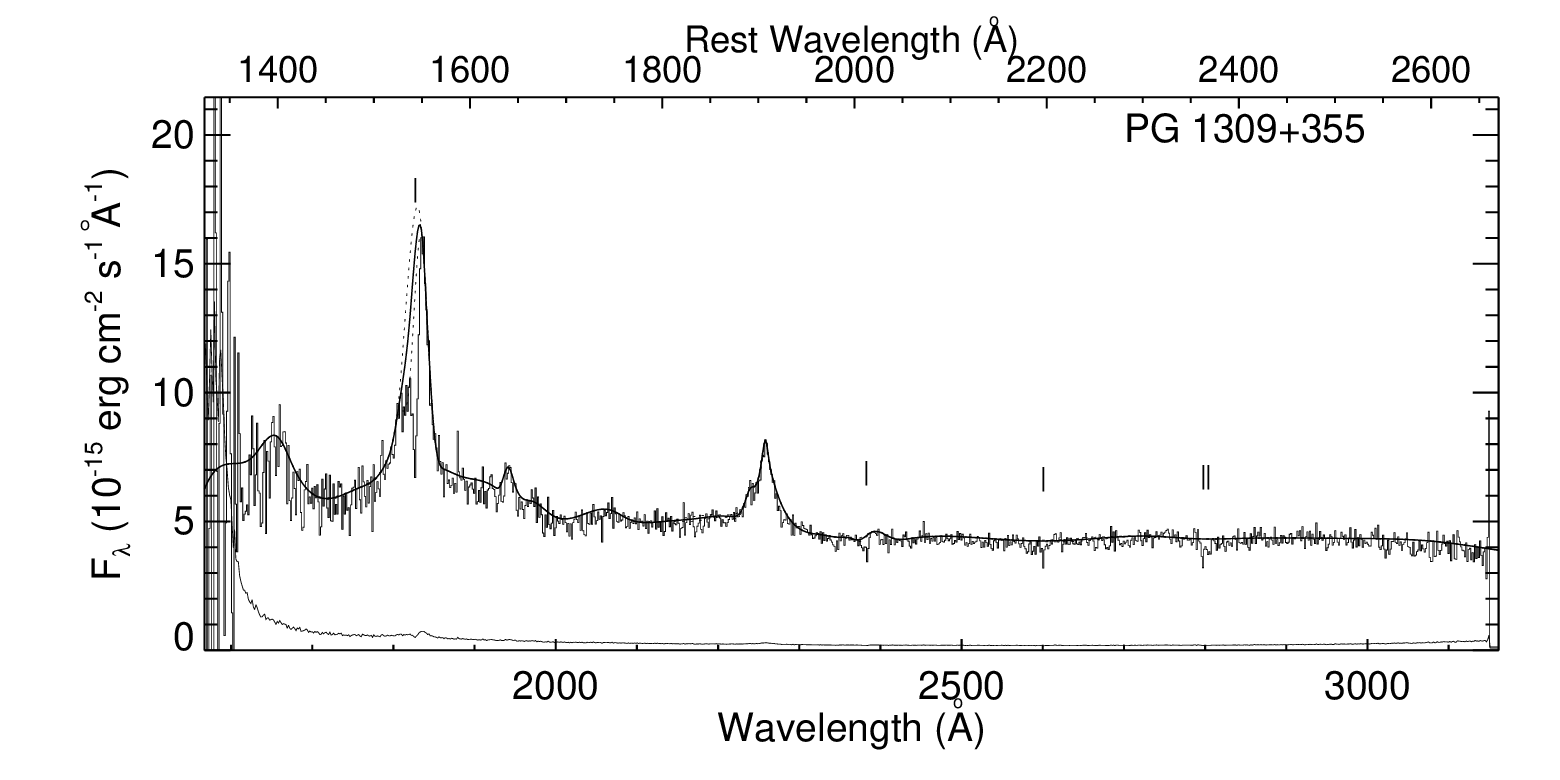}
            \includegraphics[scale=0.4,angle=90]{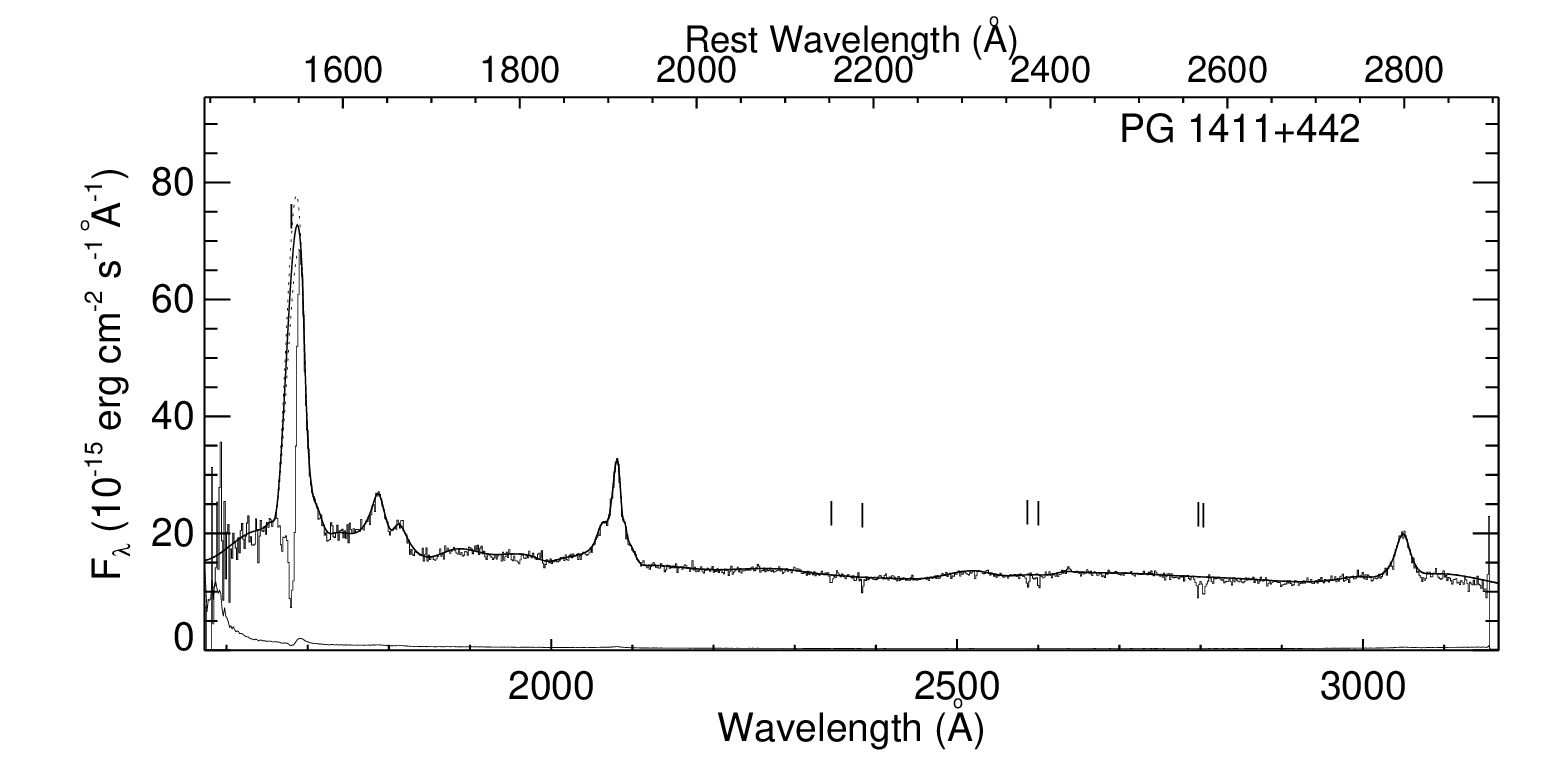}
            \includegraphics[scale=0.4,angle=90]{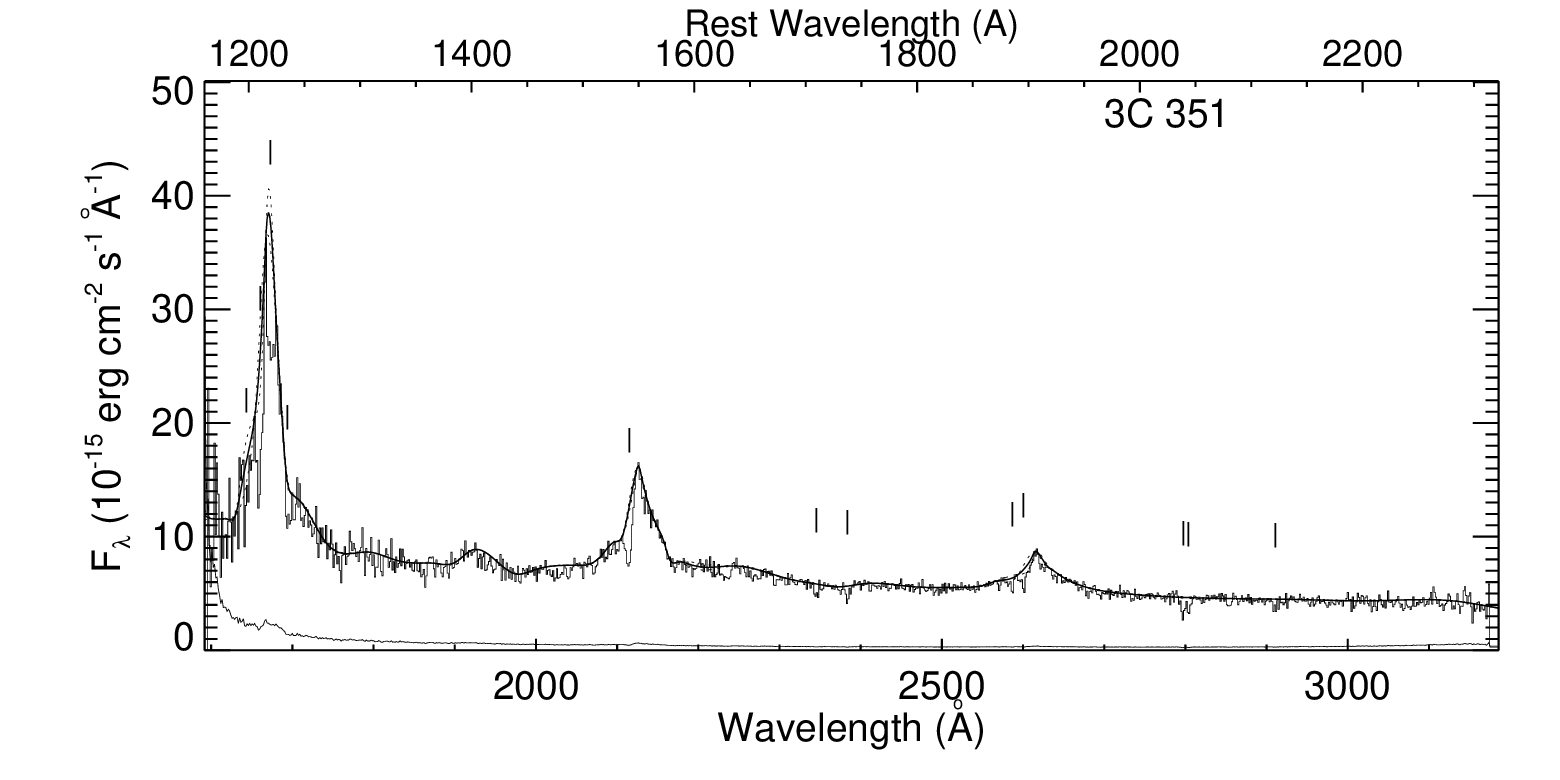}}
\bigskip\centerline{Fig.~\ref{fig_spec} {\em (continued)}}
\end{figure}
\clearpage
\begin{figure}  
\leftline{\includegraphics[scale=0.4,angle=90]{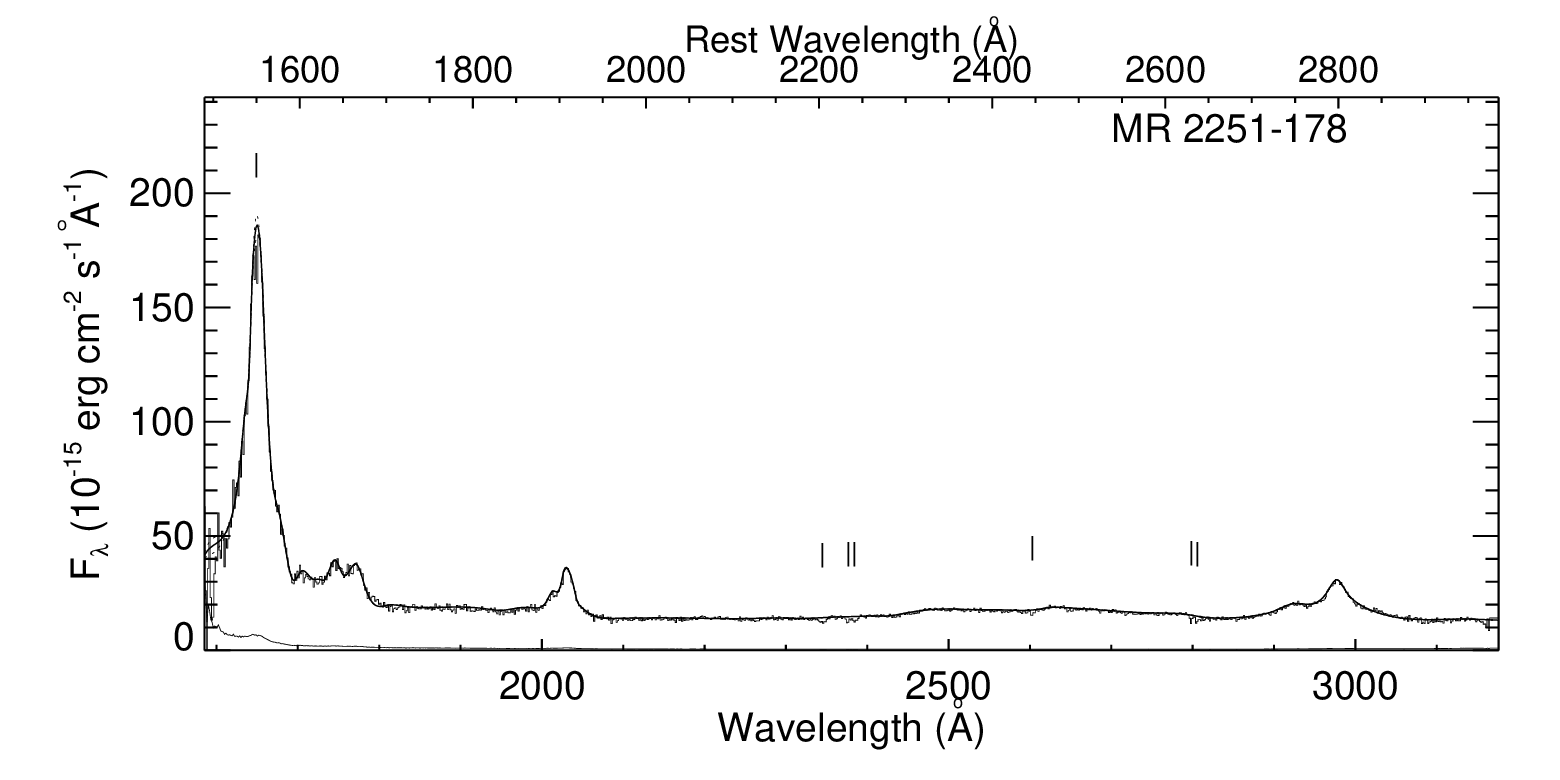}}
\leftline{\includegraphics[scale=0.4,angle=90]{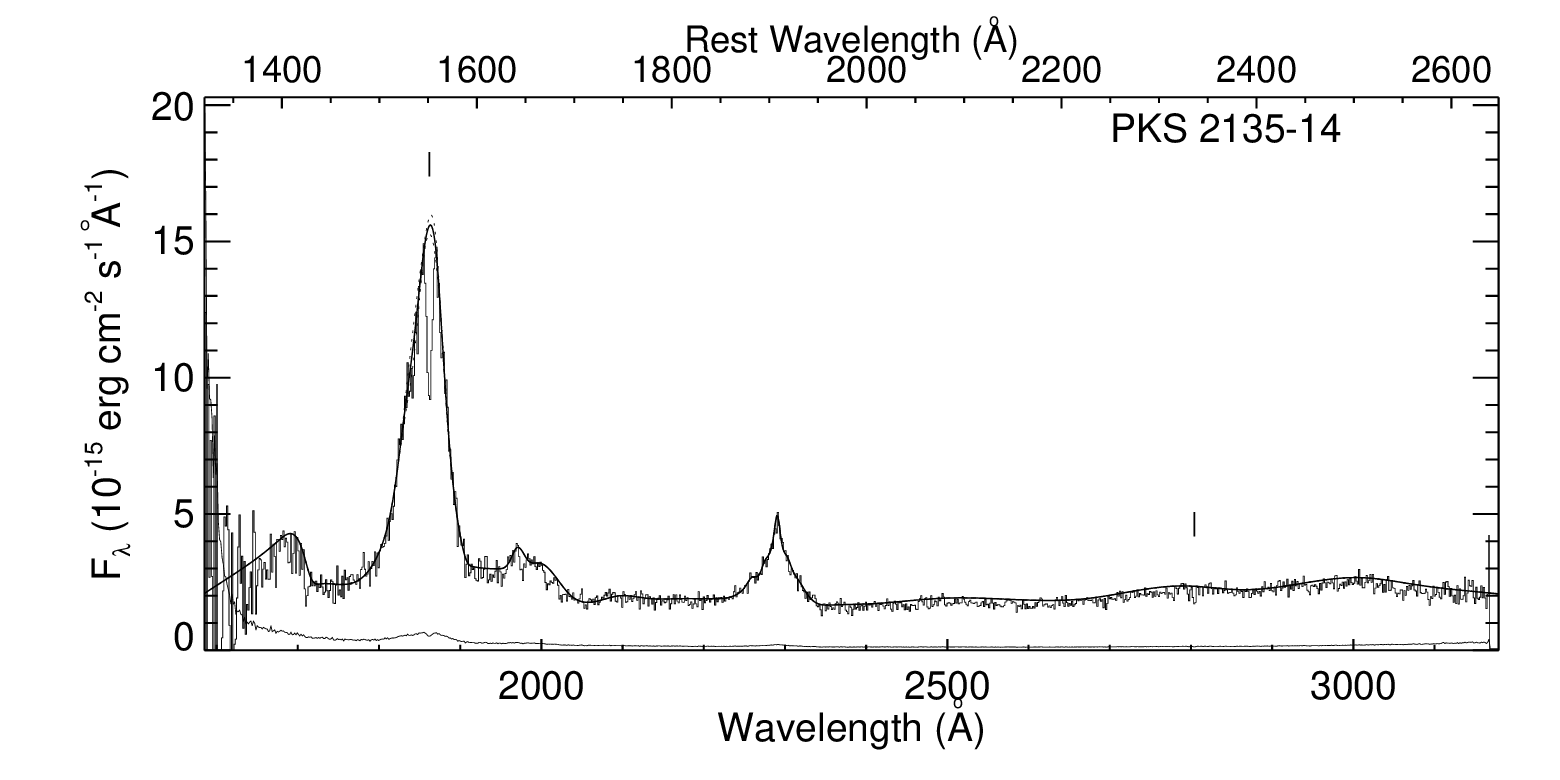}
          \includegraphics[scale=0.4,angle=90]{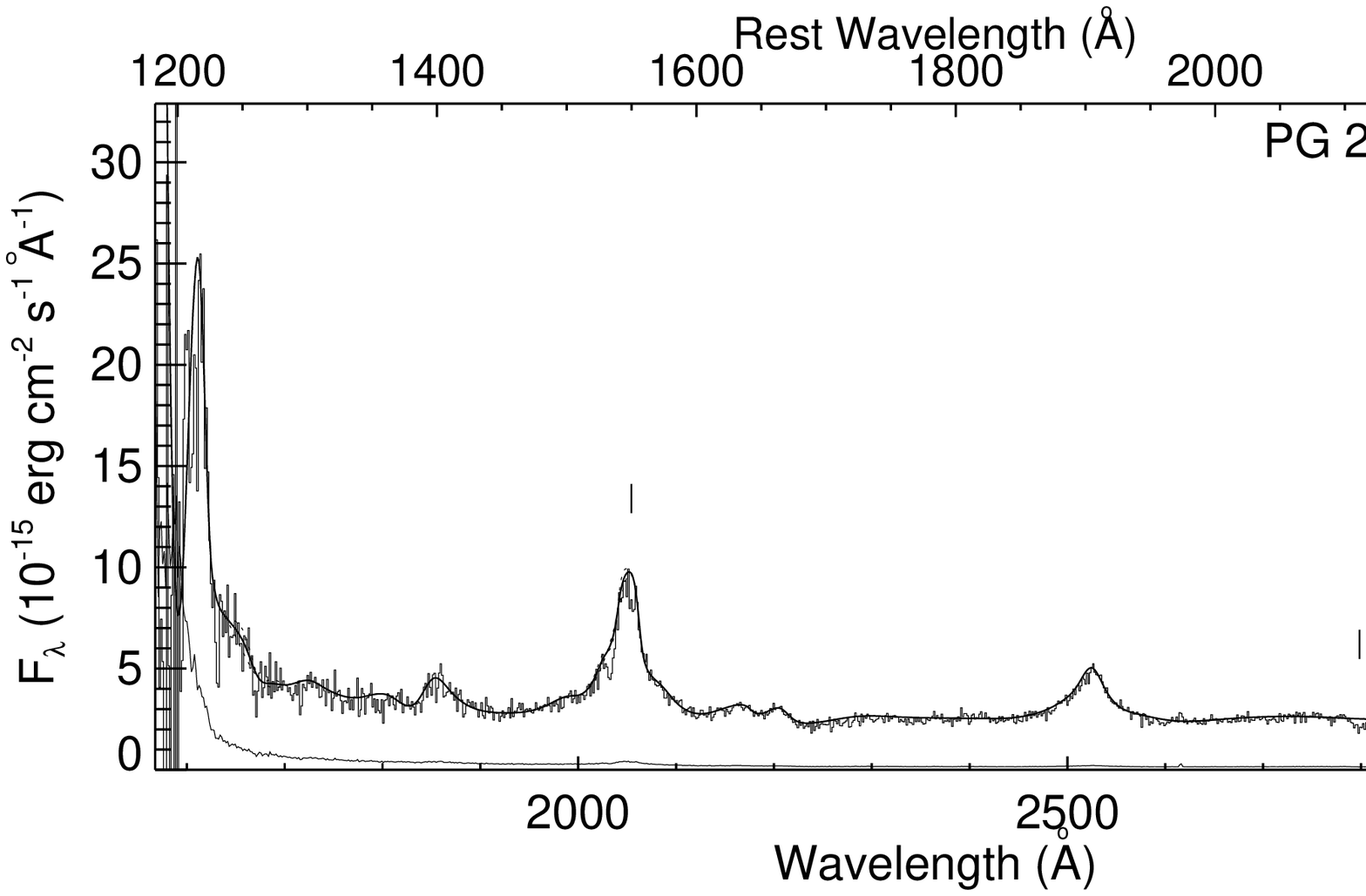}}
\bigskip\centerline{Fig.~\ref{fig_spec} {\em (continued)}}
\end{figure}
\clearpage
\begin{figure} 
\centerline{\includegraphics[scale=0.6,angle=-90]{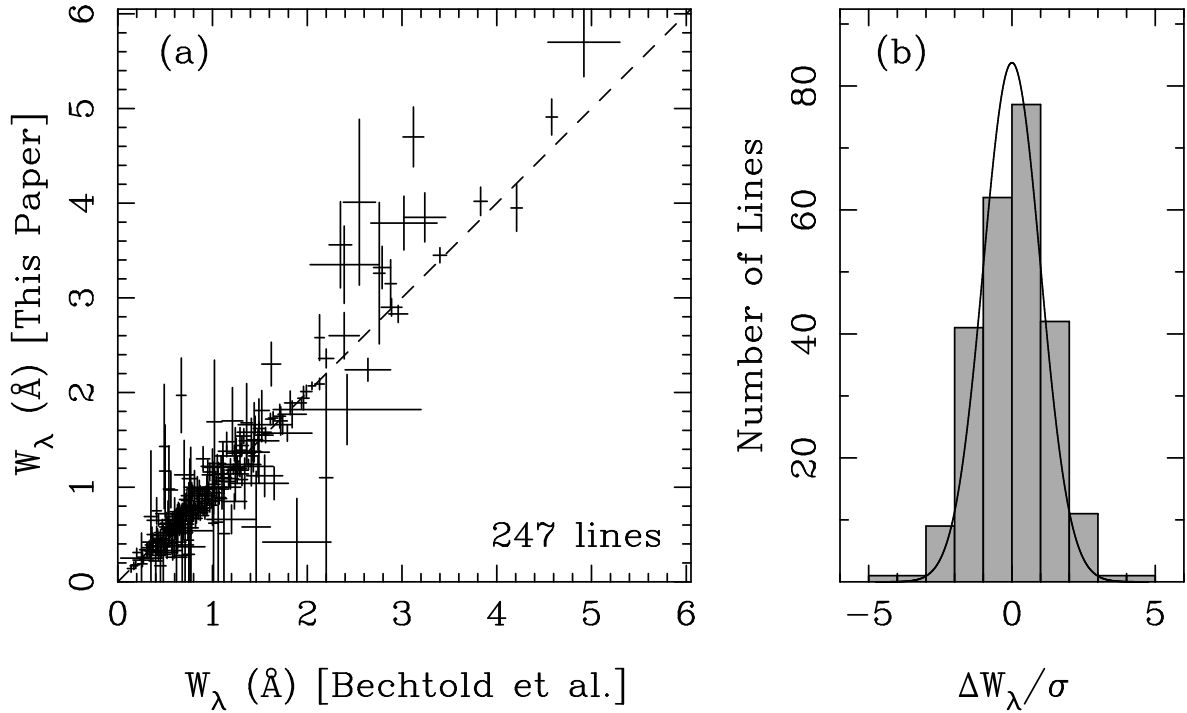}}
\caption{Comparison of the EWs of 247 lines that we have measured in
  FOS spectra with measurements of the same lines by \citet{Be02}.
  (a) Our EW measurement plotted against that of \citet{Be02}. The
  dashed line has unit slope and illustrates the good agreement
  between the two sets of measurements. In the interest of clarity,
  this plot shows only the range of observed EWs up to 6~\AA; three
  lines with $6\;{\rm \AA}<W_{\lambda}< 20\;{\rm \AA}$ are omitted.
  (b) The distribution of the normalized deviation between the two
  measurements, namely $\Delta W_{\lambda}/\sigma \equiv
  \left[W_{\lambda}({\rm theirs}) - W_{\lambda}({\rm ours}) \right]/
  (\sigma^2_{\rm theirs} + \sigma^2_{\rm ours})^{1/2}$. This
  histogram, which comprises all lines, appears symmetric about zero
  and includes five outliers with $|\Delta W_{\lambda}/\sigma|>3$ (one
  is out of the range of the plot at $|\Delta
  W_{\lambda}/\sigma|\approx 8$). For comparison, we overplot as a
  solid line a Gaussian of unit standard deviation, which represents
  the {\it expected} distribution of $\Delta W_{\lambda}/\sigma$.
\label{fig_comp}}
\end{figure}
%
\clearpage
\begin{figure} 
\centerline{\includegraphics[scale=0.65,angle=-90]{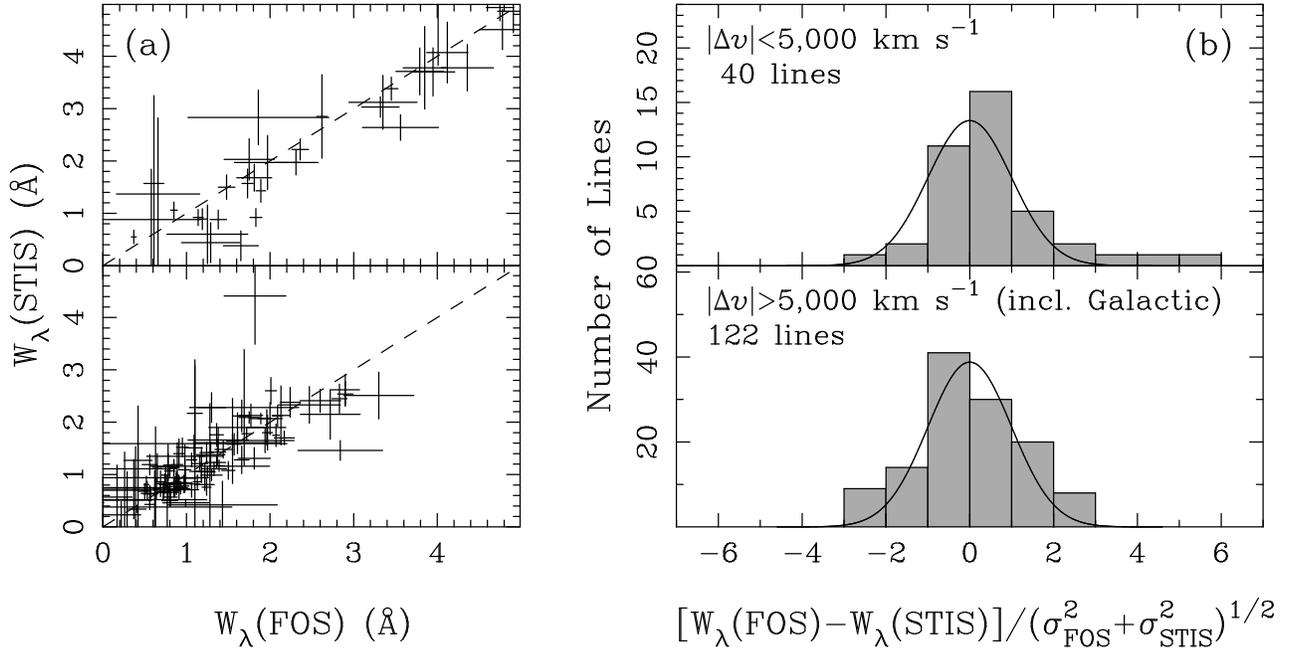}}
\caption{Distribution of differences between the EWs of the same lines
  as measured by the STIS and the FOS in two velocity bins relative to
  the quasar redshift: $|\Delta v|<5,000$\kms\ (associated lines; top
  panel) and $|\Delta v|>5,000$\kms\ (including Galactic and
  unidentified lines; bottom panel). See \S\ref{S_obs} of the text for
  an explanation of this division. Included in this figure are the two
  lines with STIS detections and FOS upper limits, bringing the total
  number of lines to 40 associated and 122 non-associated.  (a) A plot
  of $W_{\lambda}({\rm STIS})~ vs ~W_{\lambda}({\rm FOS})$ restricted
  to EW up to 5~\AA\ for clarity. The dashed line in each panel has
  unit slope.  (b) The distribution of the EW differences normalized
  by the uncertainty, $\left[W_{\lambda}({\rm FOS}) - W_{\lambda}({\rm
  STIS}) \right]/ (\sigma^2_{\rm FOS} + \sigma^2_{\rm
  STIS})^{1/2}$. The error bars ($\sigma_{\rm STIS}$ and $\sigma_{\rm
  STIS}$) include contributions from photon noise uncertainties and
  continuum placement uncertainties. The solid line overplotted on
  each histogram is a Gaussian of unit standard deviation, which is
  the distribution expected {\it a priori}, if the normalized EW
  differences are a result of measurement errors only.
\label{fig_scatter}}
\end{figure}
%
\begin{figure}  
\centerline{\includegraphics[scale=0.6]{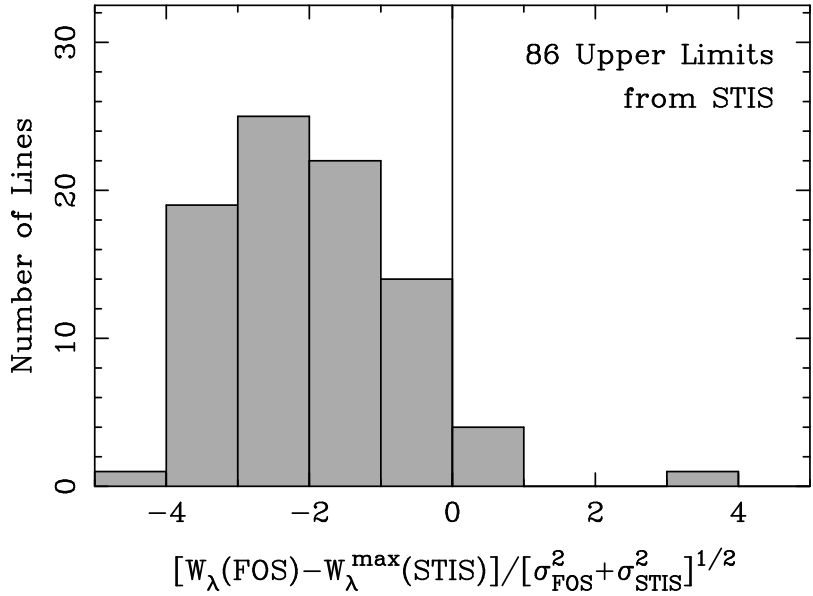}}
\caption{Distribution of the normalized EW differences of lines
  measured in FOS spectra but not detected in STIS spectra. This is
  analogous to the plot in the Figure~\ref{fig_scatter}b, but with
  $W_{\lambda}({\rm STIS})$ replaced by the 5$\sigma$ upper limit
  $W^{\rm max}_{\lambda}({\rm STIS})$. The 3.3$\sigma$ outlier on the
  far right is a Ly$\alpha$ line in PG~1718+481 at $\Delta
  v=+469$\kms.
\label{fig_limits}}
\end{figure}
%
\clearpage
\begin{figure}  
\hbox{
\hskip -0.3truein
\includegraphics[angle=-90,scale=0.37]{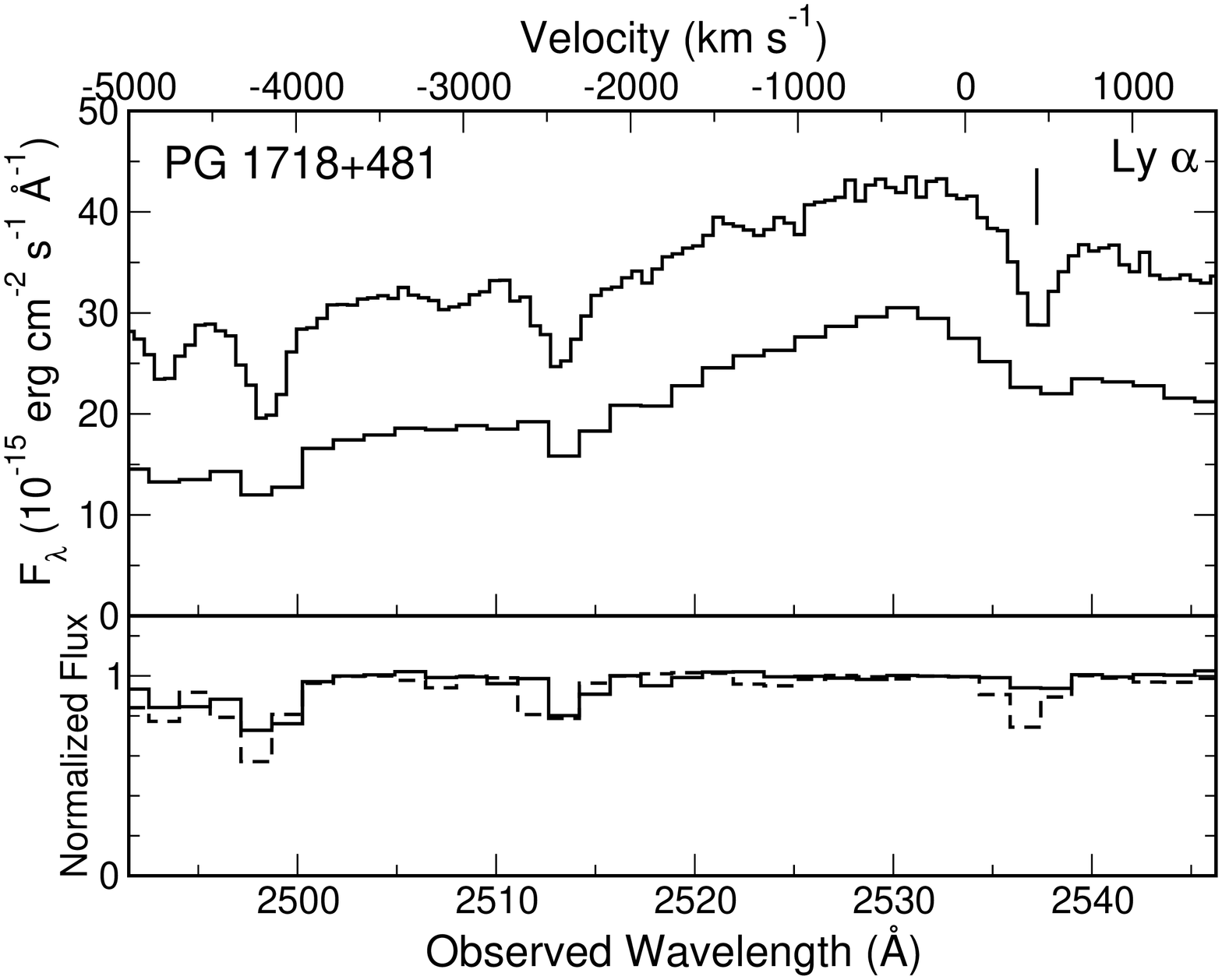}
\hskip -0.3truein
\includegraphics[angle=-90,scale=0.37]{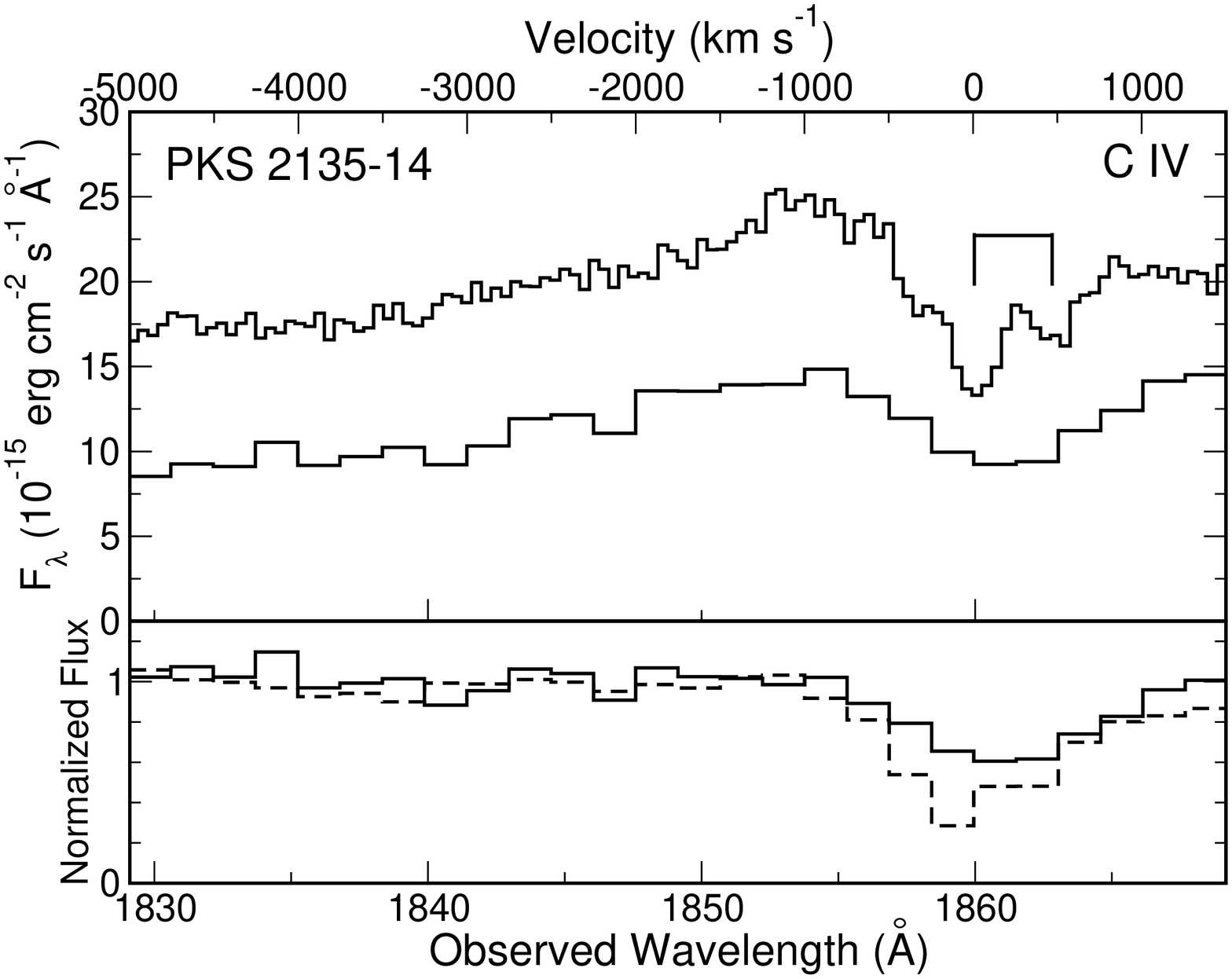}
}\vskip -0.15truein 
\hbox{
\hskip -0.3truein
\includegraphics[angle=-90,scale=0.37]{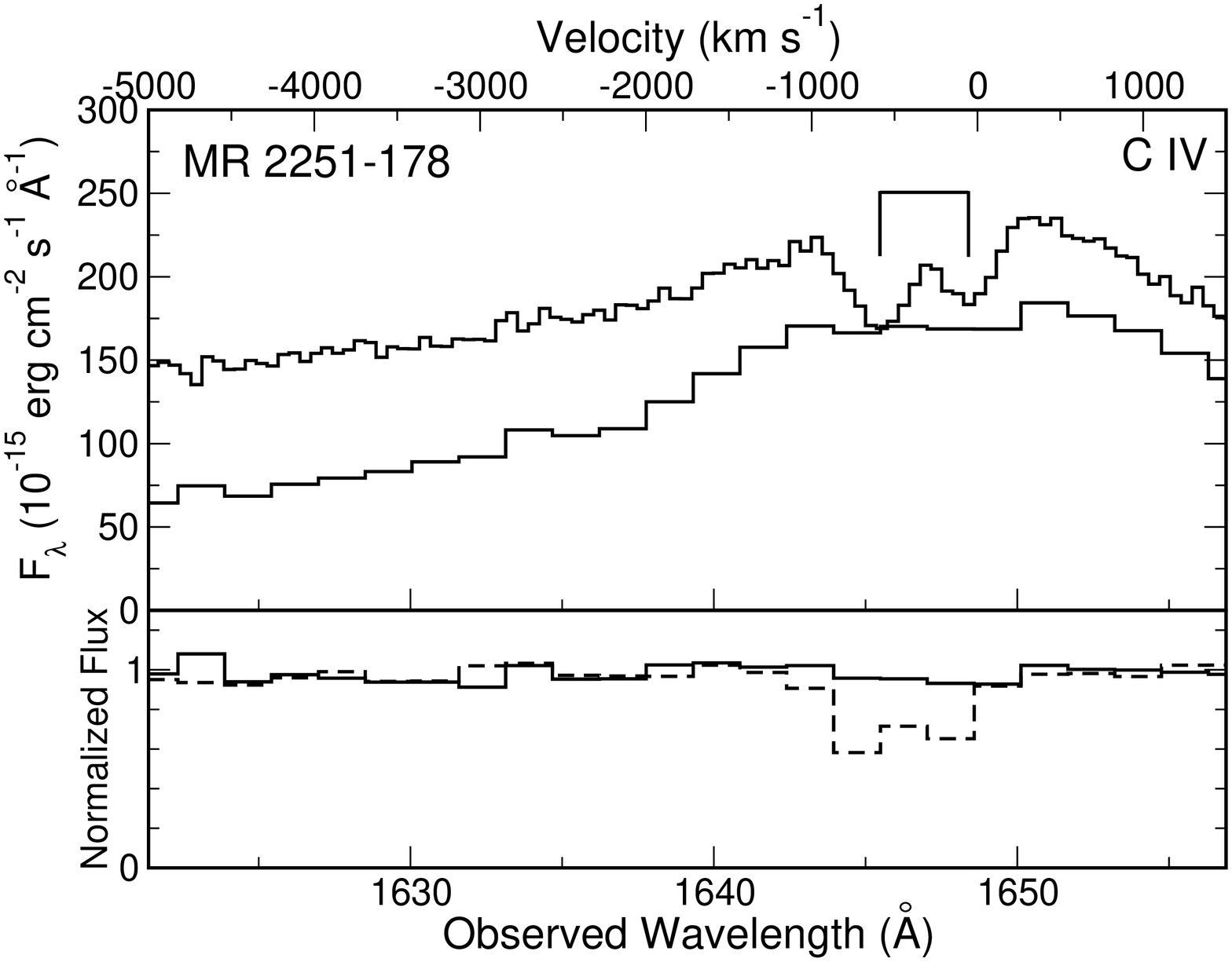}
\hskip -0.3truein
\includegraphics[angle=-90,scale=0.37]{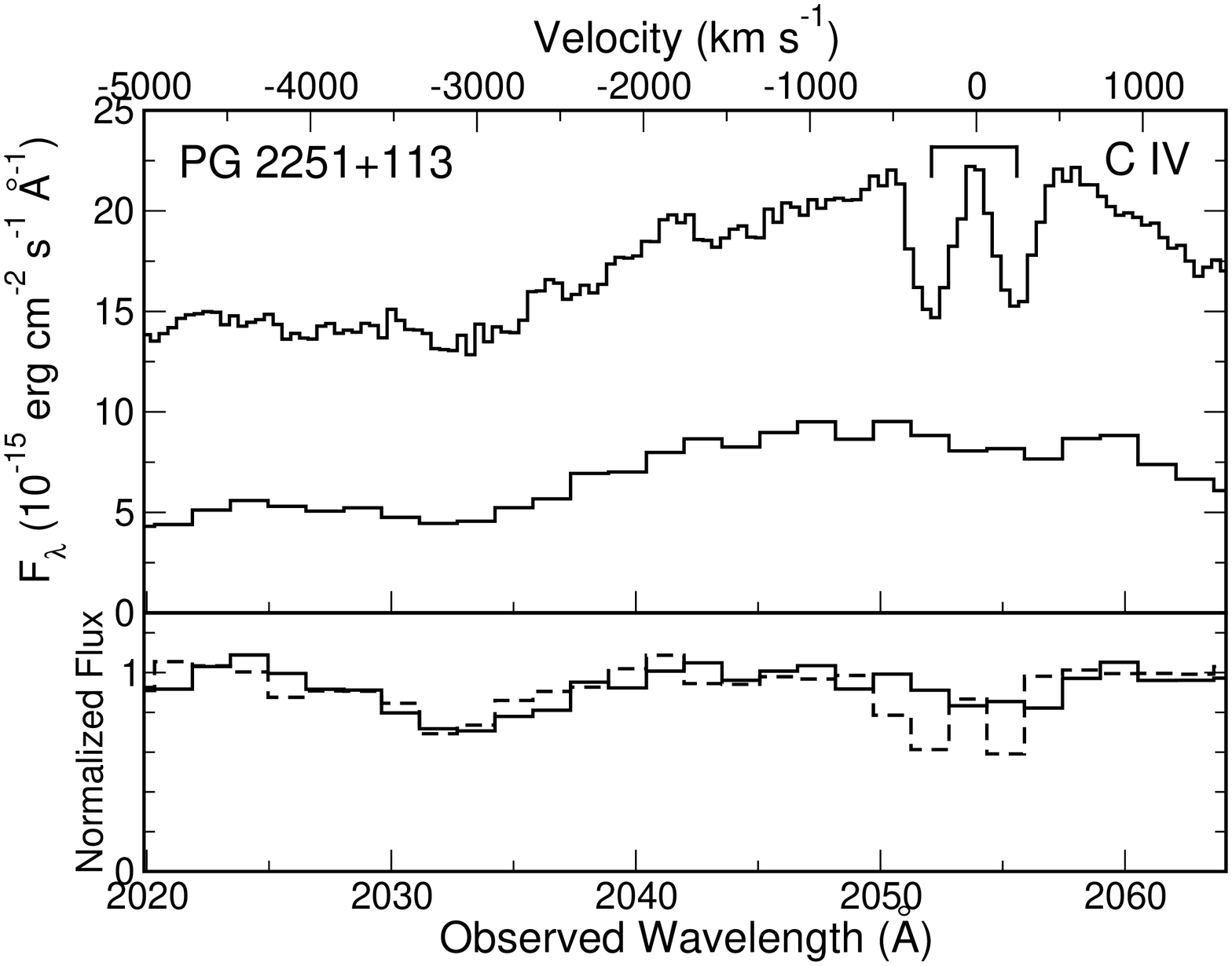}
}
\caption{ Visual illustration of the the variable NALs found in our
  variability survey. The top panel in each set shows STIS and FOS
  spectra overplotted. The FOS spectrum can be distinguished by its
  higher sampling rate and by the fact that it has been shifted
  upwards for clarity. The absorption lines of interest are identified
  by tick-marks.  The bottom panel in each set shows the
  continuum-normalized spectra from the FOS (dashed line) and STIS
  (solid line).  The FOS spectrum has been smoothed to the STIS
  resolution and resampled to the STIS wavelength scale to provide a
  fair comparison.  The wavelength scales of the normalized spectra
  were aligned by comparing the observed wavelengths of Galactic,
  interstellar lines.\label{fig_overp}}
\end{figure}
\clearpage
\begin{figure}  
\centerline{\includegraphics[scale=0.5,angle=-90]{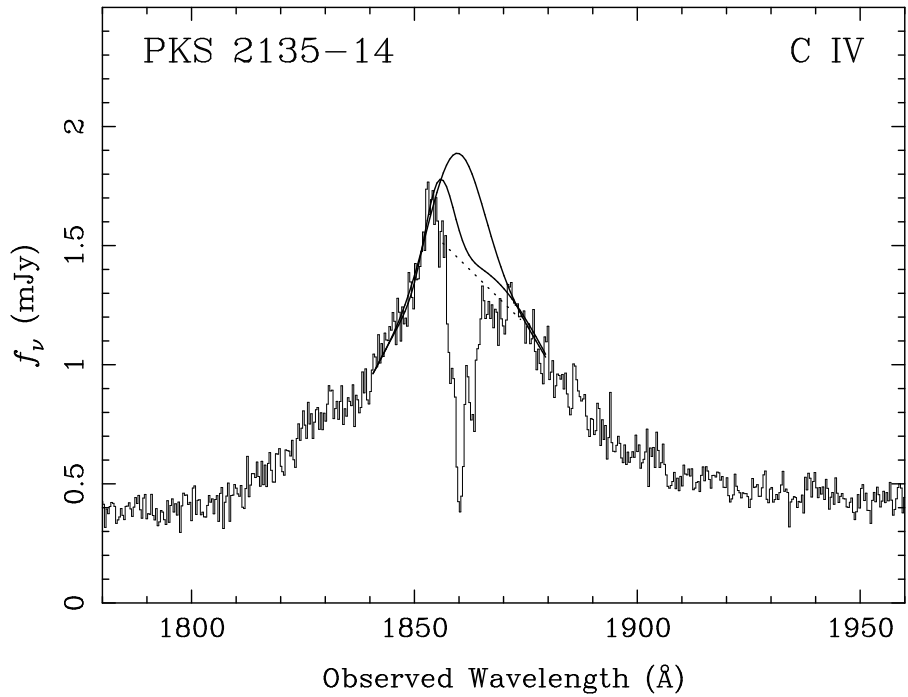}}
\caption{The \ion{C}{4} emission-line spectrum of PKS~2135--14 with
  two extreme {\it effective} continuum fits superposed as smooth,
  solid lines. The dotted line is a linear interpolation over the main
  absorption trough; it approximates the fit used by \citet{Be02} to
  measure the EW of the same absorption line.
\label{fig_pksfits}}
\end{figure}
\clearpage
\begin{figure}
\centerline{\includegraphics[scale=1.1]{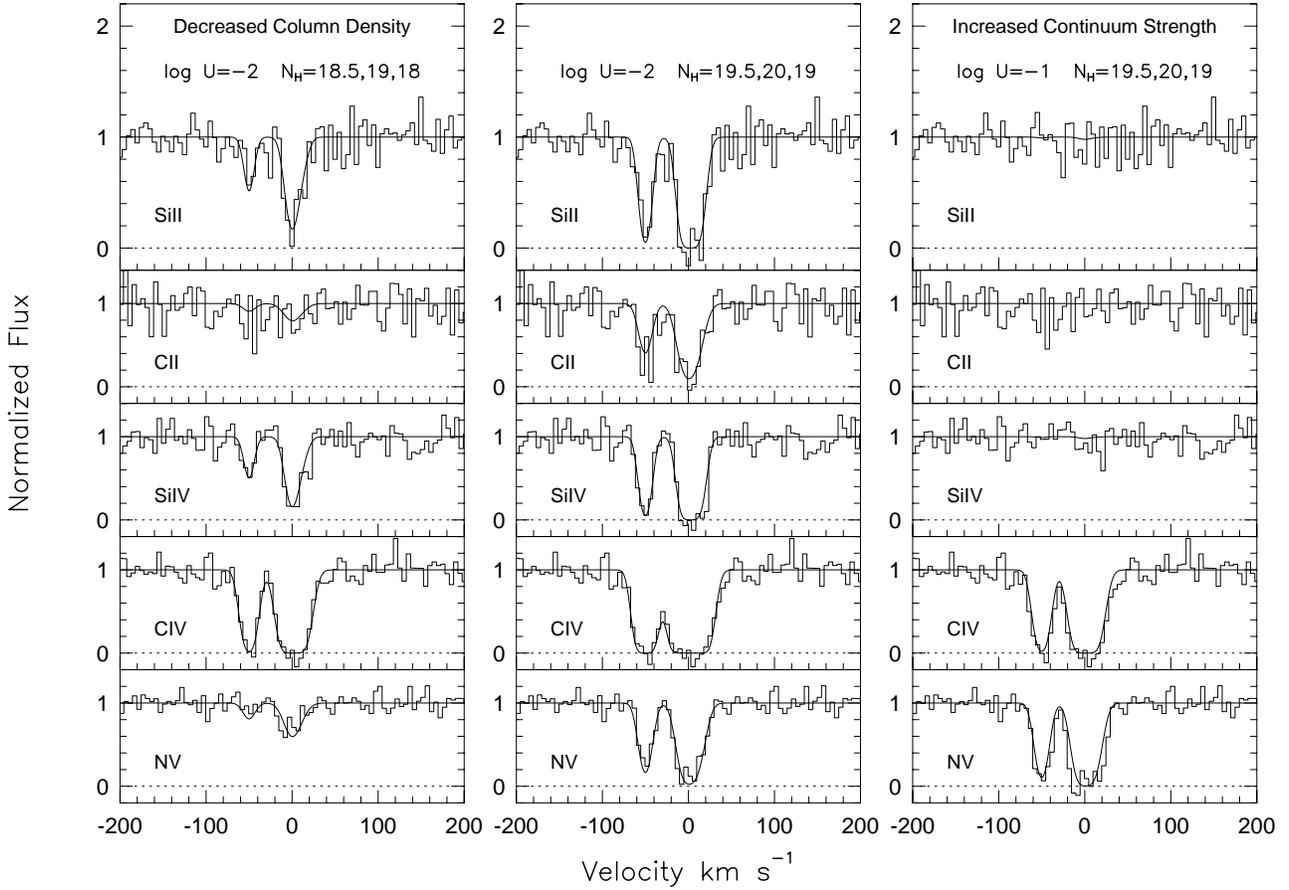}}
\caption{ Results of simulations demonstrating the effect of bulk
  motion and changing continuum strength on a series of atomic
  transitions of a hypothetical NAL system. The details of the
  simulations are described in \S\ref{S_fut} of the text. The middle
  set of panels shows the initial state of the absorber. The set of
  panels on the left show the effect of decreasing the column density
  by an order of magnitude. The set of panels on the right show the
  effect of increasing the ionization parameter by an order of
  magnitude.
\label{fig_synth}}
\end{figure}
\end{document}

%% file: tab2.tex

\def\top{$\Sigma$}
\def\mid{{\tiny 1}}
\def\bot{{\tiny 2}}
\def\ext{{\tiny 3}}
\def\db{\tablenotemark{a}}
\def\tn#1{\tablenotemark{#1}}
\def\ul#1{\underline{\underline{#1}}}

\begin{deluxetable}{rrrrcrrrrrrlrrlr}
\tabletypesize{\footnotesize}
\tablewidth{0pc}
\tablecaption{Line Identifications and Equivalent Width Measurements\label{tab_var}}
\rotate
\tablehead{
\multicolumn{4}{c}{STIS} & & \multicolumn{4}{c}{FOS} & \multicolumn{2}{c}{} & \multicolumn{4}{c}{Line ID} & \\
\noalign{\vskip -4pt}
\multicolumn{4}{c}{\hrulefill} & & \multicolumn{4}{c}{\hrulefill} & \multicolumn{2}{c}{} & \multicolumn{4}{c}{\hrulefill} & \\
\colhead{$\lambda_{obs}$} & \colhead{$W_{\lambda}$} & \colhead{$\sigma_{ph}$} & \colhead{$\sigma_{cont}$} & \colhead{Blend}&
\colhead{$\lambda_{obs}$} & \colhead{$W_{\lambda}$} & \colhead{$\sigma_{ph}$} & \colhead{$\sigma_{cont}$} & 
\colhead{\hbox to 3.0em{\hfil$\Delta W_{\lambda}$}} & \colhead{} & 
\colhead{} & \colhead{$\lambda_{rest}$} & \colhead{$\Delta\lambda$} & \colhead{} & \colhead{$\Delta v$} \\
\colhead{\hbox to 2.0em{\hfil (\AA)\hfil}} & \colhead{\hbox to 2.0em{\hfil (\AA)\hfil}} & \colhead{\hbox to 2.0em{\hfil (\AA)\hfil}} & \colhead{\hbox to 2.0em{\hfil (\AA)\hfil}} & 
\colhead{Flag} &
\colhead{\hbox to 2.0em{\hfil (\AA)\hfil}} & \colhead{\hbox to 2.0em{\hfil (\AA)\hfil}} & \colhead{\hbox to 2.0em{\hfil (\AA)\hfil}} & \colhead{\hbox to 2.0em{\hfil (\AA)\hfil}} &
\colhead{\hbox to 3.0em{\hfil\hfil (\AA)\hfil}} & \colhead{\hbox to 3.0em{\hfil \large $\frac{\Delta W_{\lambda}}{\sigma}$\hfil}} &
\colhead{\hbox to 2.0em{\hfil Ion\hfil}} & 
\colhead{\hbox to 2.0em{\hfil (\AA)\hfil}} & \colhead{\hbox to 2.0em{\hfil (\AA)\hfil}} & 
\colhead{\hbox to 2.0em{z$_{abs}$}} & \colhead{\hbox to 3 em{(km s$^{-1}$)}} \\
\colhead{(1)} & \colhead{(2)} & \colhead{(3)} & \colhead{(4)} & \colhead{(5)} &
\colhead{(6)} & \colhead{(7)} & \colhead{(8)} & \colhead{(9)} &
\colhead{(10)} & \colhead{(11)} &
\colhead{(12)} & \colhead{(13)} & \colhead{(14)} & \colhead{(15)} & \colhead{(16)}
}
\startdata
\multicolumn{15}{c}{EX 0302--223  \quad ($z_{\rm e} = 1.409)$}\\
\noalign{\vskip 3pt \hrule \vskip 3pt}
 2158.66 &    0.84 &  0.22 &  0.84 &      & \multicolumn4{c}{out of range}  &    \dots &     \dots &       \dots     &   \dots    &   \dots &   \dots &   \dots \\
 2169.36 &    1.07 &  0.23 &  1.07 &      & \multicolumn4{c}{out of range}  &    \dots &     \dots &       \dots     &   \dots    &   \dots &   \dots &   \dots \\
 2185.40 &    1.83 &  0.32 &  1.83 &      & \multicolumn4{c}{out of range}  &    \dots &     \dots & \ion{H}{1}      &  937.80    &    2.14 & 1.32806 & $-$9945 \\
 2202.38 &    1.58 &  0.26 &  1.58 &      & \multicolumn4{c}{out of range}  &    \dots &     \dots &       \dots     &   \dots    &   \dots &   \dots &   \dots \\
 2212.46 &    2.39 &  0.27 &  2.39 &      & \multicolumn4{c}{out of range}  &    \dots &     \dots & \ion{H}{1}      &  949.74    &    1.40 & 1.32806 &$-$10048 \\
 2254.02 &    1.50 &  0.23 &  0.05 &      & 2253.75 &  1.48 &  0.10 &  0.00 &  $-$0.01 &   $-$0.06 & \ion{H}{1}      &  937.80    & $-$1.88 & 1.40523 &  $-$719 \\
 2266.48 &    2.45 &  0.22 &  0.17 &      & 2264.09 &  2.83 &  0.09 &  0.00 &     0.38 &      1.32 & \ion{H}{1}      &  972.54    & $-$0.03 & 1.32806 &$-$10224 \\
\enddata
\tablecomments{See \S\ref{S_obs} of the text for a detailed explanation of the organization of this table and meaning of the columns.}
\tablenotetext{a}{Mean wavelength of doublet; adopted when the doublet is unresolved in both the FOS and STIS spectra.}
\tablenotetext{b}{The \lion{Si}{4}{1393} line at $z_a=1.009$ in the spectrum of EX~0302--223 is contaminated by Galactic \lion{Mg}{2}{2803}.}
\tablenotetext{c}{The Ly$\alpha$ line at $z_a=1.125$ in the spectrum of QSO~0957+561A is contaminated by Galactic \lion{Fe}{2}{2586}.}
\tablenotetext{d}{In the case of PG~1241+176, the Galactic \lion{Fe}{2}{2344} line and the \lion{O}{6}{1032} line at $z_a=1.2735$ are blended 
                  in both the STIS and FOS spectra.}
\tablenotetext{e}{In the case of PG~1718+481, the \lion{O}{6}{1037} and \lion{Fe}{2}{2344} lines appear to be blended with neighboring unidentified lines 
                  in both the STIS and FOS spectra.}
\tablenotetext{f}{The \ion{C}{4} doublet at $z_a=0.2014$ in the spectrum of PKS~2135--14 is slightly contaminated by the Galactic \ion{Al}{3} $\lambda$1863 line.}
\tablenotetext{g}{The observed lines at $\lambda$2376 and $\lambda$2384 in the spectrum of MRC~2251--178 are unidentified. However, they are contaminated by 
                  the Galactic \ion{Fe}{2} $\lambda$2344 and $\lambda$2374, respectively.}
\end{deluxetable}